\documentclass{article}

\usepackage{amsmath}
\usepackage{amsfonts}
\usepackage{graphicx}
\usepackage{caption}
\usepackage{subcaption}

\usepackage{authblk}
\usepackage{amssymb}

\usepackage{epstopdf}
\usepackage{epsfig}
\usepackage{hyperref}

\usepackage{color}

\date{\today}

\begin{document}

\title{Quasinormal modes of Einstein-Gauss-Bonnet-dilaton black holes}

\author[1]{Jose Luis Bl\'azquez-Salcedo}
\author[1,2]{Fech Scen Khoo}
\author[1]{Jutta Kunz}
\vspace{0.5truecm}
\affil[1]{Institut f\"ur  Physik, Universit\"at Oldenburg, Postfach 2503,
D-26111 Oldenburg, Germany}
\affil[2]{Van Swinderen Institute for Particle Physics and Gravity,
University of Groningen, Nijenborgh 4, 9747 AG Groningen, The Netherlands}

\vspace{0.5truecm}

\vspace{0.5truecm}

\vspace{0.5truecm}
\date{
\today}

\maketitle

\begin{abstract}
We study quasinormal modes of static Einstein-Gauss-Bonnet-dilaton black holes.
Both axial and polar perturbations are considered and studied from $l=0$ to 
$l=3$. We emphasize the difference in the spectrum between the Schwarzschild 
solutions and dilatonic black holes.  At large Gauss-Bonnet coupling constant
a small secondary branch of black holes is present, when the dilaton coupling 
is sufficiently strong. The modes of the primary branch can differ from the 
Schwarzschild modes up to $10\%$. The secondary branch is unstable and possesses
long-lived modes. We address the possible effects of these modes on future 
observations of gravitational waves emitted during the ringdown phase of 
astrophysical black holes. %
\end{abstract}

\section{Introduction}

The recent direct detections of gravitational waves by the LIGO/Virgo 
Collaboration \cite{PhysRevLett.116.061102,PhysRevLett.116.241103,PhysRevLett.118.221101} 
have fueled the interest in the study of black holes and neutron stars 
in astrophysical systems. The observational data are compatible with 
black hole mergers in general relativity (GR). The first observed event
resulted in the formation of a black hole of $62 M_{\odot}$, the 
second event resulted in a $21 M_{\odot}$ black hole, and the very recent third event resulted in a $49 M_{\odot}$ black hole. 
In the three cases, the estimated energy radiated in gravitational waves was 
about $5\%$ of the final black hole mass. However, although the detections 
from LIGO appear to be in good agreement with the predictions from GR, 
there is still plenty of room for alternative gravitational theories 
\cite{DeLaurentis:2016jfs,Konoplya:2016pmh,PhysRevLett.116.171101,Cardoso:2016oxy,Abedi:2016hgu,Nakano:2017fvh,Sibandze:2017jbi}. 
The study of black hole collisions in alternative theories of gravity, 
and, in particular, the study of the ringdown phase after the merger phase, 
can in principle be used to constrain those theories when compared with (future) 
observations.

In contrast to neutron stars, black holes are free from the extra complications
introduced by high density states of matter. 
Gravitational waves radiated after the merger of black holes depend only on 
much simpler parameters, independently of how the gravitational waves 
have been triggered.
The ringdown is determined by the three basic parameters of the final black 
hole, namely its mass, charge and angular momentum.
In particular, the parameters of the final black hole produced after the merger 
can be inferred from the damped sinusoidal waves produced during the 
ringdown phase.
The characteristic frequencies of these gravitational waves are given by 
the spectrum of quasinormal modes (QNM) of the black hole. 

Although quasinormal modes contain the generic information on the source 
of the gravitational waves, the ringdown phase can differ 
in alternative theories of gravity from GR. 
The deviations in the QNM spectrum from GR predictions can be used 
to test alternative gravitational theories in the strong field regime. 
Additionally, a QNM analysis allows one to study the stability of black holes 
under linear mode perturbations.

From an observational perspective, the GR description of gravity 
may be questioned since it leads to an evolution of the Universe, 
that is currently dominated by the unknown constituents
dark matter and dark energy. 
From the theoretical point of view
an alternative theory of gravity is required 
in order to consistently formulate a quantum theory of the fundamental 
interactions. Here, one of the promising contenders for a new theoretical
framework is string theory.
In string theory, the Einstein-Hilbert term of GR will receive 
higher order corrections. 
For instance, the low-energy effective action
as obtained from heterotic string theory
will include a certain quadratic curvature term,
the Gauss-Bonnet term, together with a coupling to a dilaton field
\cite{Gross:1986mw,Metsaev:1987zx}.
In particular, after compactification and truncation of 
extra gauge fields in heterotic string theory, an effective theory
emerges containing both ingredients.

Thus inspired from string theory 
\cite{Mignemi:1992nt,Mignemi:1993ce,Kanti:1995vq}
Einstein-Gauss-Bonnet-dilaton (EGBd) theory 
is a gravitational theory which extends the Einstein-Hilbert action 
to include a dynamical scalar degree of freedom which is coupled 
to the Gauss-Bonnet invariant.
In four dimensions, the Gauss-Bonnet invariant is topological, 
i.e., only a total derivative in the action,
which would not modify the Einstein field equations by itself.
But its non-minimal coupling to the dilaton field 
circumvents this.
Moreover,
despite the higher order curvature terms, the field equations of the theory 
remain second order in the derivatives.
This ensures that no propagating ghost degree of freedom 
(in the sense of Ostrogradsky \cite{Os}) arises. 
Note, that the Gauss-Bonnet-dilaton term of the action also 
appears as an ingredient of a more general scalar-tensor theory 
(as the `Ringo' term in the Fab-Four), which is directly related to a 
subsector of Horndeski gravity \cite{Berti:2015itd,Maselli:2016gxk}.

Thus EGBd theory 
represents an interesting effective theory of gravity 
that can be tested by observations.
It is viable on astrophysical scales \cite{Berti:2015itd}.
In particular, it is in accordance with solar system observations
and it allows for new effects and deviations from GR in strong fields.
It features two coupling constants, the Gauss-Bonnet (GB) coupling
constant $\alpha$ and the dilaton coupling constant $\gamma$.
There exist already some constraints imposed on the theory
from astronomical observations \cite{Blazquez-Salcedo:2016yka}, 
which have been derived for the (heterotic string theory) value $\gamma=1$.
Measurement of the Shapiro time delay from the Cassini mission imposes 
a mild upper limit on the GB coupling constant, 
$\alpha = 10^{26} \text{cm}^2$ \cite{Bertotti:2003rm}. 
A stronger constraint is obtained from the observation of the low-mass 
X-ray binary A0620-00, based on the orbital decay rate in a black hole 
\cite{0004-637X-710-2-1127}. 
This observation constrains the maximum value of the possible dilaton charge, 
and hence the coupling constant $\sqrt{\alpha} = 3.8 \times 10^{5} \text{cm}$ 
\cite{Yagi:2012gp}.

Black hole solutions of EGBd theory could not be found in analytical form.
Thus, static EGBd black hole solutions were first studied perturbatively
\cite{Mignemi:1992nt,Mignemi:1993ce} and then numerically 
\cite{Kanti:1995vq,Torii:1996yi}. Recently, the numerical solution has been analytically approximated in \cite{Kokkotas:2017ymc} using a continued fraction expansion.
These EGBd black holes carry nontrivial dilaton fields outside their horizon.
The presence of the Gauss-Bonnet-dilaton terms in the generalized
Einstein equations lead to an ``effective'' energy-momentum tensor,
allowing for negative ``effective'' energy densities and
thus an evasion of the classical ``no-scalar-hair'' theorem
\cite{Kanti:1995vq}.
Furthermore, the early studies \cite{Kanti:1995vq} of static EGBd black
holes showed already that the domain of existence of these solutions 
is bounded by a maximum value of the GB coupling parameter $\alpha$.
Depending on the value of the dilaton coupling $\gamma$,
close to the maximum of $\alpha$ a short secondary branch of black hole solutions
can arise \cite{Torii:1996yi,Guo:2008hf}. 

Rotating EGBd black holes were studied in the slow rotation approximation in 
\cite{PhysRevD.79.084031,Pani:2011gy,Ayzenberg:2014aka,Maselli:2015tta}, 
and in the fast rotating case in 
\cite{PhysRevLett.106.151104,Kleihaus:2014lba,Kleihaus:2015aje}. 
Analogously to the static case, the domain of existence is bounded. 
The effect of rotation is to cause an effective reduction in the range 
of the coupling constant, where rotating black hole solutions exist.
Interestingly, the rapidly rotating EGBd black holes can slightly
exceed the Kerr bound for the angular momentum, and their
quadrupole moments and moments of inertia can deviate significantly
from the respective Kerr values.
However, their shadows are very close to the shadows of Kerr black holes
\cite{Cunha:2016wzk}, and an analogous observation
holds for their X-ray reflection spectrum
\cite{Zhang:2017unx}.

Stability of the static EGBd black holes was studied in 
\cite{Kanti:1997br,PhysRevD.58.084004}, showing stability
with respect to radial fluctuations for black holes on the
primary branch, but revealing instability on the short
secondary branch.
First studies of axial and polar gravitational perturbations 
were done in \cite{PhysRevD.79.084031,Blazquez-Salcedo:2016enn},
indicating linear mode stability of the black holes
on the primary branch.
A similar conclusion was reached in \cite{Ayzenberg:2013wua},
where the polar sector was analyzed 
in the high-frequency, geometric optics approximation.

In this paper we continue the analysis of the quasinormal modes of 
static EGBd black holes performed in \cite{Blazquez-Salcedo:2016enn}.
We present results for quasinormal modes from $l=0$ to $l=3$, 
in the axial and polar perturbations.
We focus on perturbations on the full numerical background. 
This also allows us to conduct a closer inspection of the quasinormal modes 
close to the maximal value of the GB coupling constant, 
where the effect of the dilaton field is strong.
The quasinormal mode analysis reveals a complicated structure 
in the spectrum close to this value, 
which might have observational implications.
Our results from the analysis of quasinormal modes
support both the linear mode stability of the static EGBd black holes
on the primary branch, as well as the instability of
the solutions belonging to the short secondary branches, 
which can be present for large values of the coupling $\alpha$
(depending on the value of the dilaton coupling $\gamma$)
\cite{Guo:2008hf,PhysRevD.58.084004}.
Finally, using results from the recent observation of gravitational 
waves GW151226 \cite{PhysRevLett.116.241103},
we discuss how EGBd theory could show its effects 
on the ringdown frequencies in 
black hole mergers. Moreover we address the bounds of the coupling constants.

\section{Static black holes in Einstein-Gauss-Bonnet-dilaton theory} \label{sec_theory}

\subsection{Theoretical framework}

In four dimensions, the Einstein-Gauss-Bonnet-dilaton action is given by
\cite{Kanti:1995vq}

\begin{equation}
S_{\text{EGBd}}(g,\Phi) = \frac{1}{16\pi } \int d^4x 
\sqrt{-g}\biggl(R - \frac{1}{2}\partial_{\mu}\Phi\,\partial^{\mu}\Phi + \frac{1}{4}\alpha e^{\gamma\Phi}R_{\text{GB}}^2 \biggr)
\label{action}
\end{equation}
in natural units $c=1=G$, 
where the first term in the action is the standard Einstein-Hilbert term, 
followed by the kinetic term for the dilaton field $\Phi$. 
$R_{\text{GB}}^2$ is the Gauss-Bonnet term
\begin{equation}
R_{\text{GB}}^2 = R_{\mu\nu\rho\sigma}R^{\mu\nu\rho\sigma} -4R_{\mu\nu}R^{\mu\nu} + R^2~,
\end{equation}
which is coupled to the dilaton via the GB coupling constant $\alpha$ 
and the dilaton coupling constant $\gamma$.

The field equations for the metric in this theory are
\begin{equation}
\label{metfeqn}
G_{\mu\nu} = \frac{1}{2}\biggl(\partial_{\mu}\Phi \,\partial_{\nu}\Phi -\frac{1}{2}g_{\mu\nu}\partial_{\rho}\Phi\, \partial^{\rho}\Phi \biggr) - \frac{1}{4} \alpha e^{\gamma\Phi}\biggl(H_{\mu\nu} + 4(\partial^{\rho}\Phi \, \partial^{\sigma}\Phi +\partial^{\rho}\partial^{\sigma}\Phi)P_{\mu\rho\nu\sigma}  \biggr)~,
\end{equation}
where $G_{\mu\nu}$ is the usual Einstein tensor, 
and the gravitational terms on the right hand side arise 
due to the coupling to the dilaton field, where 
\begin{eqnarray}
H_{\mu\nu}=2(RR_{\mu\nu}-2R_{\mu\sigma}R^{\sigma}_{\ \nu}-2R_{\mu\rho\nu\sigma}R^{\rho\sigma}+R_{\mu\rho\sigma\lambda}R_{\nu}
^{\ \rho\sigma\lambda})-\frac{1}{2}g_{\mu\nu}R^2_{\text{GB}}~, \\
P_{\mu\nu\rho\sigma}=R_{\mu\nu\rho\sigma}+g_{\mu\sigma}R_{\rho\nu}-g_{\mu\rho}R_{\sigma\nu}+g_{\nu\rho}R_{\sigma\mu}-g_{\nu\sigma}R_{\rho\mu}
+\frac{1}{2}Rg_{\mu\rho}g_{\sigma\nu}-\frac{1}{2}Rg_{\mu\sigma}g_{\rho\nu}~.
\end{eqnarray}
$H_{\mu\nu}$ does not contribute, since this tensor vanishes
in four dimensions.
The dilaton field equation is obtained from the variation of the action 
with respect to $\Phi$
\begin{equation}
\label{dilfeqn}
\nabla^2 \Phi=\frac{1}{4}\alpha\gamma e^{\gamma\Phi}R_{\text{GB}}^2~.
\end{equation} 
From the field equations (\ref{metfeqn}) and (\ref{dilfeqn}) it is clear
that they are at most second order in the derivatives.

An interesting limit of the theory is obtained when the dilaton coupling is linear. This case can be obtained by modifying the Gauss-Bonnet term in the previous action Eq.(\ref{action}): $\frac{1}{4}\alpha e^{\gamma\Phi} R_{\text{GB}}^2 \to \frac{1}{4}\alpha \gamma\Phi R_{\text{GB}}^2$. The linear dilaton coupling can be understood as the first nontrivial term of the small $\gamma$ limit in the exponential coupling. Note that $\gamma$ then becomes trivial since it can be absorbed by redefining $\alpha \to \alpha/\gamma$ without loss of generality. 

To describe static, spherically symmetric black holes 
we consider the following ansatz for the metric 
\begin{equation}
\label{ds2}
ds^{2}=g^{(0)}_{\mu\nu}dx^{\mu}dx^{\nu}=-F(r)dt^{2} + K(r)dr^{2} + r^{2}(d\theta^{2} 
+ \sin^{2}\theta \, d\varphi^{2})~.
\end{equation} 
The metric functions $F(r)$ and $K(r)$, 
together with the dilaton field $\Phi_0(r)$ 
are functions of the radial coordinate $r$, and can
be obtained by inserting the ansatz into
the equations (\ref{metfeqn})-(\ref{dilfeqn})
and then solving the resulting system of coupled ordinary
differential equations numerically,
subject to appropriate boundary conditions.

The black hole solutions are characterized 
by the presence of an event horizon at $r=r_H$, 
where 
$$K(r) \approx \frac{1}{1-2m_1}\cdot\frac{r_H}{r-r_H}, $$ 
$$F(r) \approx f_1 (r - r_H)$$ 
with a constant coefficient $f_1$, and 
$$\Phi_0 (r) \approx \Phi_{00} + \Phi_{01} (r-r_H),$$ 
where the constant 
$m_1 = \frac{\alpha\gamma\Phi_{01}}{2\alpha\gamma\Phi_{01}
+4r_He^{\gamma\Phi_{00}}}$ 
is given in terms of the constant $\Phi_{01}$ that satisfies
\begin{equation}
\alpha \gamma r_H^2 \Phi_{01}^2 + 2 e^{\gamma \Phi_{00}}r_H^3 \Phi_{01} + 6 \alpha \gamma=0~.
\label{bcPhi01}
\end{equation} 
$\Phi_{00} := \Phi_0(r_H)$ is the dilaton value at the horizon. 
This is a quadratic equation for $\Phi_{01}$, 
which should have real solutions, so that the dilaton field is 
regular and possesses a real valued expansion near the horizon.
Consequently the radicand of the solutions for $\Phi_{01}$
should not become negative, and
this requires that regular black hole solutions satisfy the inequality
\begin{equation}
e^{2\gamma \Phi_{00}} r_H^4 -6 \alpha^2 \gamma^2 > 0~.
\label{relPhi00}
\end{equation} 
It is this inequality, which leads to the theoretical bound
on the GB coupling $\alpha$ or the product $\gamma\alpha$, when $\gamma$ is varied \cite{Kanti:1995vq}.

Asymptotically the solutions approach the Schwarzschild spacetime, 
which means $F(r) \approx 1 - 2M/r + f_2/r^2$, 
$K(r) \approx 1 + 2M/r$, and $\Phi_0(r) = Q/r$. 
Thus the mass of the black holes $M$ and the dilaton charge $Q$
can be read off asymptotically.
$f_2$ is another expansion constant.

\subsection{Static black holes}

\begin{figure}
     \centering

     \begin{subfigure}[b]{0.4\textwidth}
\includegraphics[width=50mm,scale=0.5,angle=-90]{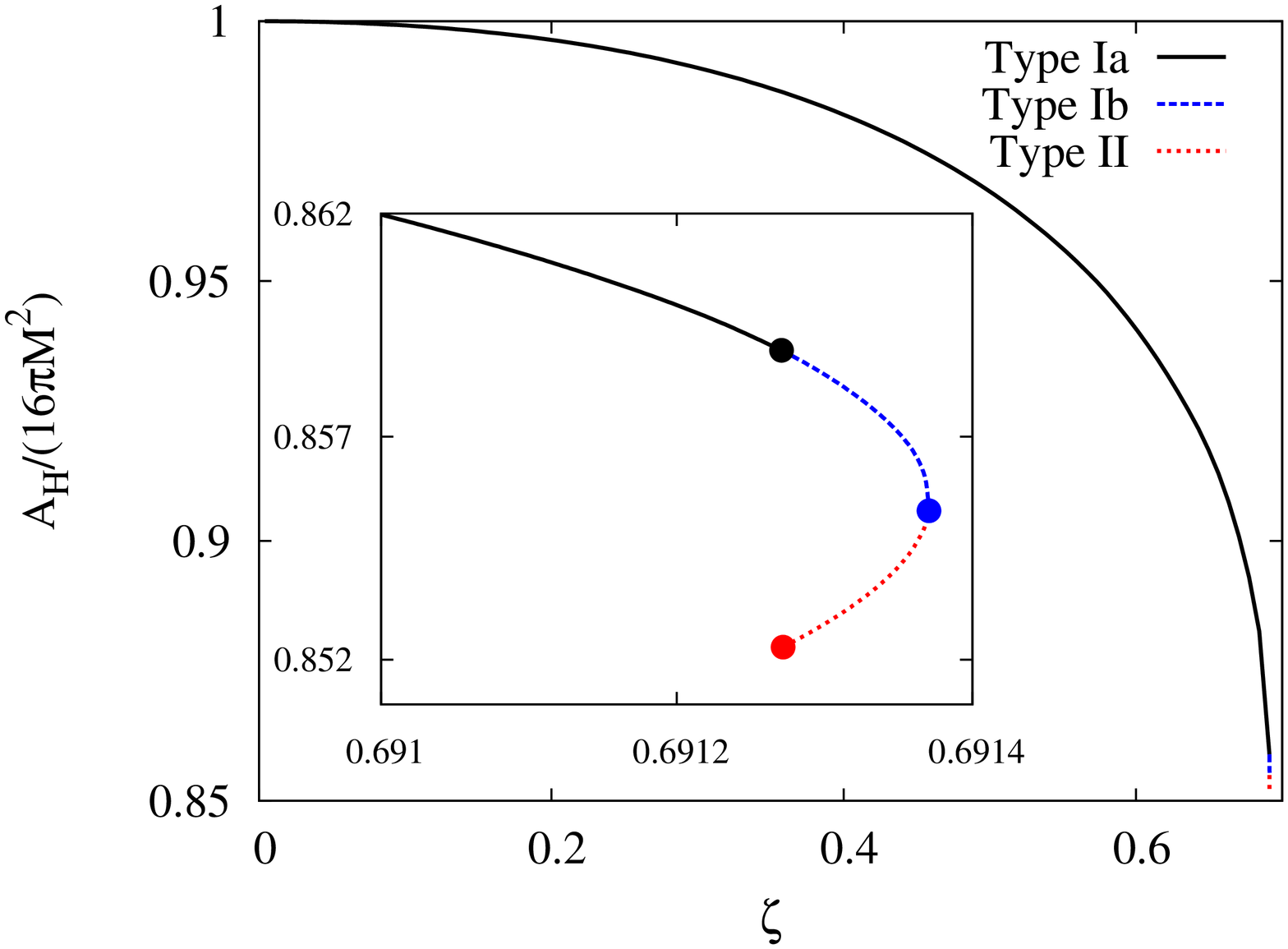}
         \caption{}
         \label{fig:area_dilaton_2branches_a}
     \end{subfigure}
     \begin{subfigure}[b]{0.4\textwidth}
\includegraphics[width=50mm,scale=0.5,angle=-90]{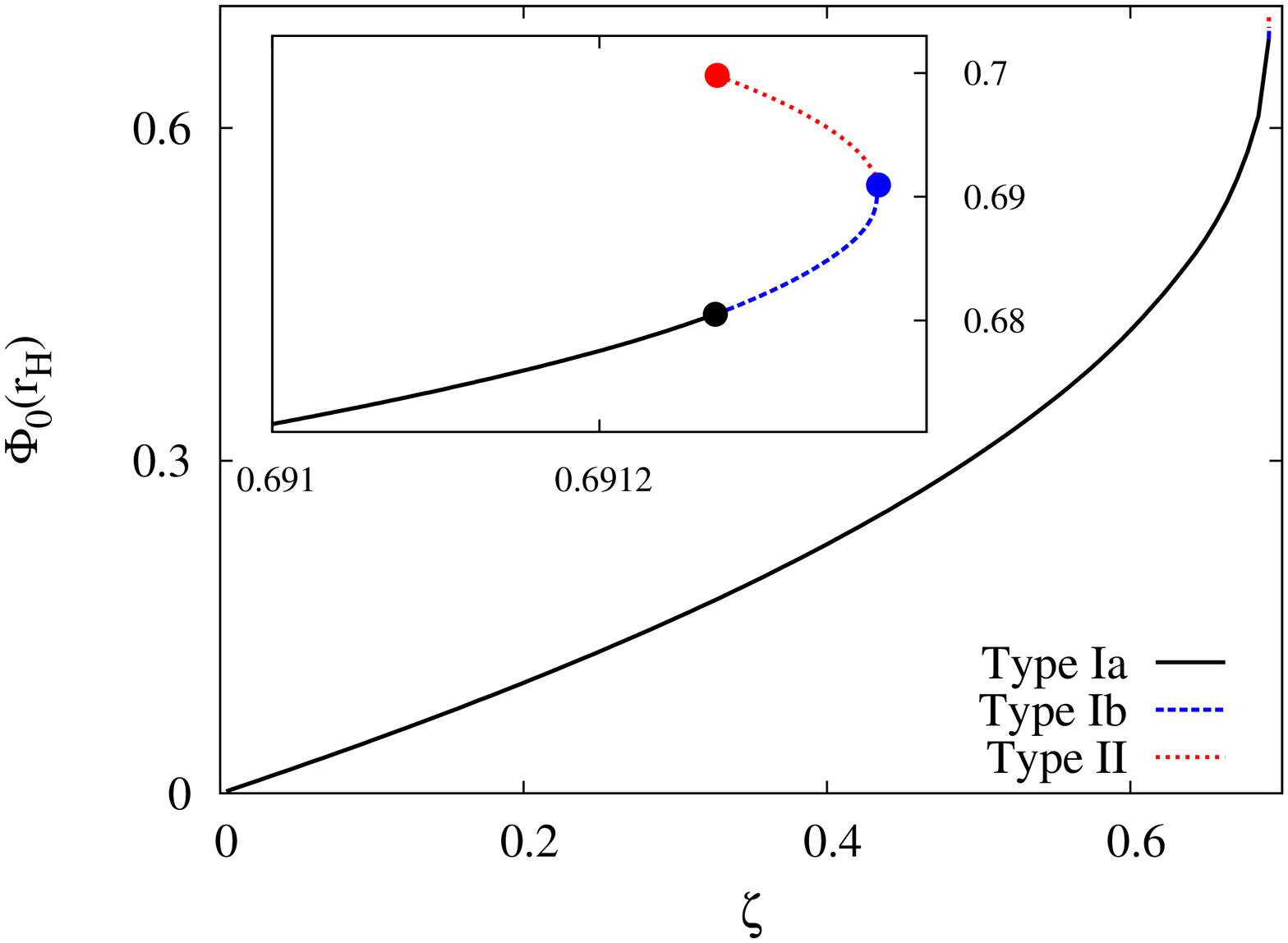}
         \caption{}
         \label{fig:area_dilaton_2branches_b}
     \end{subfigure}

     \caption{
(a) The scaled area of the horizon $A_H/16 \pi M^2$ of static EGBd black holes
versus the scaled GB coupling constant $\zeta={\alpha}/{M^2}$.
(b) The value of the dilaton field at the horizon $\Phi_0(r_H)$ 
versus the scaled GB coupling constant $\zeta$.
The insets show the solutions in the vicinity of the critical GB
coupling constant.}

         \label{fig:area_dilaton_2branches}

\end{figure}

Let us now briefly summarize the relevant properties of the static
EGBd black hole solutions.
As noted in \cite{Kanti:1995vq}, these black hole solutions possess
a nontrivial dilaton field. Since the dilaton charge $Q$ does not
represent an additional conserved charge of the system,
the black holes possess only secondary hair \cite{Hajian:2016iyp}.
Thus, for a given value of the GB coupling constant $\alpha$
(and the dilaton coupling constant $\gamma$),
the static black holes form a one parameter family of solutions,
labeled for instance by the black hole mass $M$ or by the black hole
horizon area $A_{H}$.

Another interesting property of these black holes noted in \cite{Kanti:1995vq}
is that their domain of existence is bounded.
To illustrate this we present in Fig.\ref{fig:area_dilaton_2branches}
the domain of existence of the static black holes for the value of the
dilaton coupling constant $\gamma=1$, the value most explored
in the literature so far.
In Fig.\ref{fig:area_dilaton_2branches_a} we exhibit 
the scaled area of the black hole horizon $A_H/16 \pi M^2$ 
of static EGBd black holes as a function of the scaled GB
coupling constant $\zeta=\alpha/M^2$.
(The normalization is chosen such that the Schwarzschild black holes
at $\zeta=0$ possess the value $A_H/16 \pi M^2=1$.) 
Fig.\ref{fig:area_dilaton_2branches_b} represents a similar plot
for the value of the dilaton field at the horizon $\Phi_0(r_H)$.

\begin{figure}
     \centering

     \begin{subfigure}[b]{0.4\textwidth}
\includegraphics[width=50mm,scale=0.5,angle=-90]{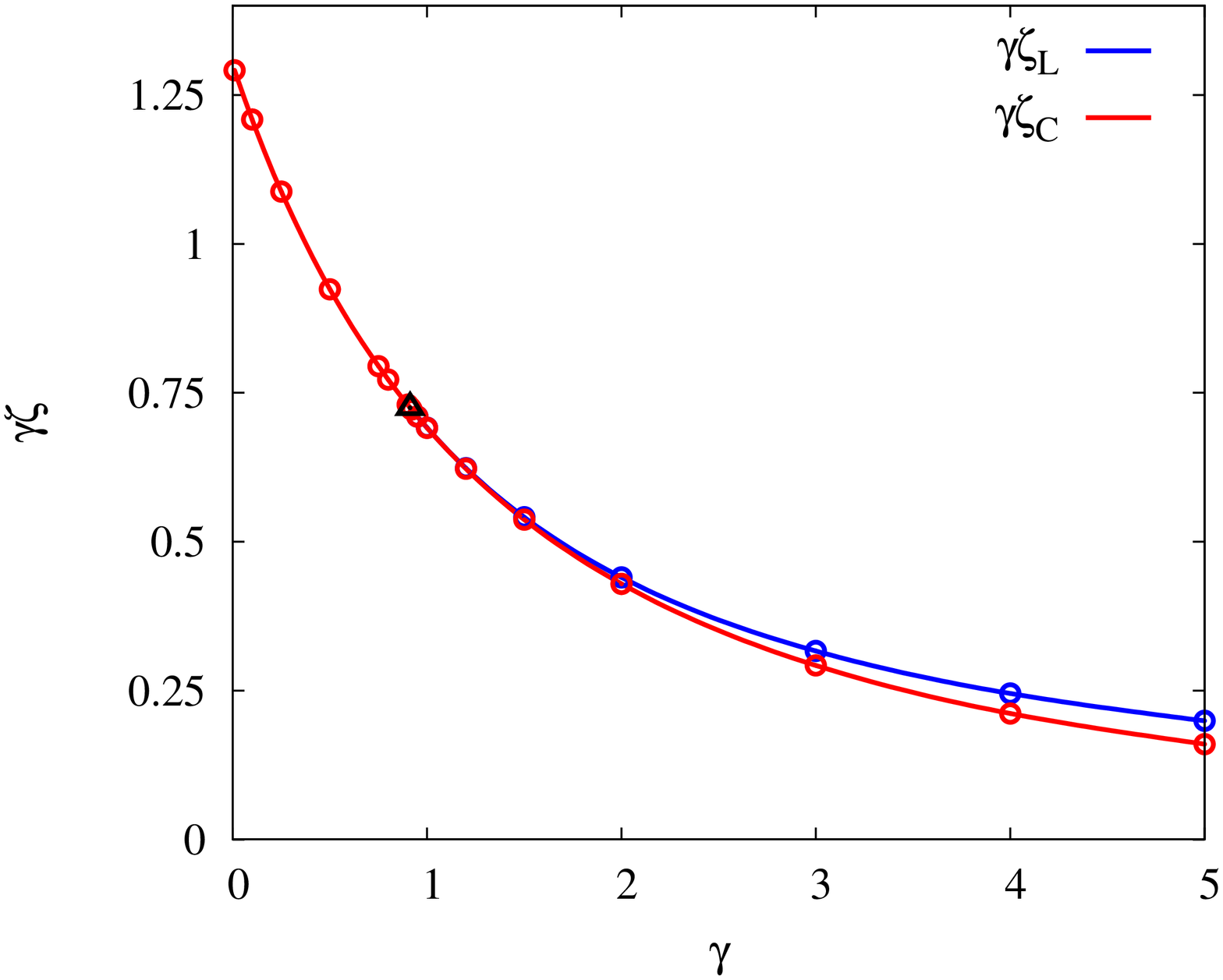}
         \caption{}
         \label{fig:gamma_alpha_a}
     \end{subfigure}
     \begin{subfigure}[b]{0.4\textwidth}
\includegraphics[width=50mm,scale=0.5,angle=-90]{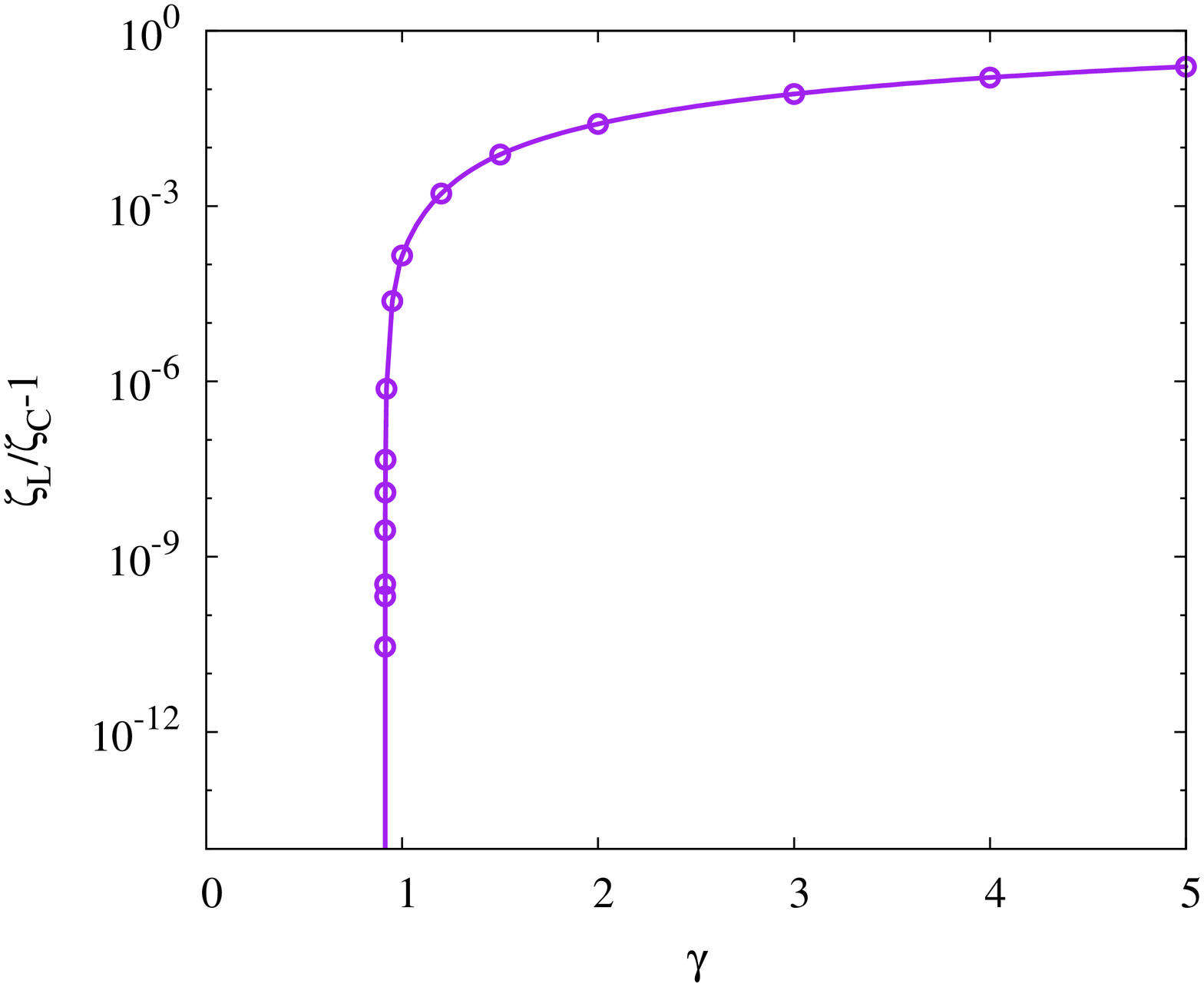}
         \caption{}
         \label{fig:gamma_alpha_b}
     \end{subfigure}

     \caption{(a) The scaled maximal value $\zeta_L$ and the 
critical value $\zeta_C$ of the GB coupling constant (multiplied by $\gamma$) versus 
the dilaton coupling constant $\gamma$ for static EGBd black holes. The triangle marks the point where the second branch ceases to exist ($\gamma=0.9130$).
(b) The deviation of the ratio $\zeta_L/\zeta_C$ from one
versus the dilaton coupling constant $\gamma$.}

         \label{fig:gamma_alpha}

\end{figure}

Starting from the Schwarzschild value, 
the area decreases as the scaled GB constant $\zeta$ rises
together with the value of the dilaton field at the horizon.
However, for $\gamma=1$, 
the branch of solutions extends only up to a 
maximal value of the scaled GB coupling, $\zeta_L = 0.691372$ 
(marked by the blue dot in Fig.\ref{fig:area_dilaton_2branches}). 
In fact, no regular black hole solutions exist for $\zeta > \zeta_L$. 
Instead there arises a secondary branch of black hole solutions,
residing between $\zeta_C = 0.69127$ and $\zeta_L$ 
\cite{Kanti:1995vq,PhysRevD.79.084031}. 

The main or primary branch of EGBd solutions thus extends from the
Schwarzschild black holes ($\zeta=0$) to the maximal value 
of the GB coupling $\zeta_L$. The black holes on this branch are named
type I black holes, and fall into two subtypes, which join 
at $\zeta_C$ (black dot).
The type Ia black holes (black line)
represent unique solutions,
while the type Ib black holes (blue line) are no longer unique, 
since for a given value of $\zeta$ in the range
$\zeta_C \le \zeta < \zeta_L$ 
there are always two distinct static black hole solutions with the same mass.
However, the two solutions can be distinguished by their differing
horizon properties and dilaton charge. 
The type I black holes are regular on and outside their horizon
for every value of $\zeta$, where they exist.

The primary branch ends at $\zeta_L$ (blue dot),
where it merges with the secondary type II branch. 
Note that precisely at the merger point $\zeta_L$ uniqueness is recovered
and only a single black hole solution is found. 
The secondary branch of type II black holes extends backwards from 
$\zeta_L$ to $\zeta_C$ (red dot).
At $\zeta_C$ a critical configuration is encountered,
where the horizon expansion 
saturates the reality condition Eq.(\ref{relPhi00}) for the dilaton field.
However, this critical configuration is no longer regular
\cite{Torii:1996yi,Guo:2008hf}.
The figures illustrate, 
that type II solutions possess a smaller horizon area and a larger value 
of dilaton field at the horizon than type I solutions. 
In general, the black hole horizon area decreases when
the dilaton field value gets larger.
That is, black holes with a larger horizon 
have a weaker dilaton field at the horizon.

This branch structure is not only found for $\gamma=1$, 
but it occurs as well for other values of the dilaton coupling constant
\cite{Guo:2008hf,Cunha:2016wzk}.
In Fig.\ref{fig:gamma_alpha}
we illustrate the dependence of the maximal value $\zeta_L$ and the
critical value $\zeta_C$ on the dilaton coupling constant $\gamma$.
Starting from $\gamma=1$, an increase of $\gamma$ leads to a rapid
decrease of both $\zeta_L$ and $\zeta_C$, and thus a rapid decrease
of the domain of existence of static EGBd black hole solutions.
At the same time, as the primary branch decreases in length,
the length of the secondary branch increases,
as seen in Fig.\ref{fig:gamma_alpha_a}.
On the other hand, when starting from $\gamma=1$ and decreasing $\gamma$,
$\zeta_L$ and $\zeta_C$ rise rapidly, while their difference
diminishes fast. The latter is illustrated in 
Fig.\ref{fig:gamma_alpha_b}, which reveals that the secondary branch
ceases to exist close to $\gamma=0.9130$. 

\begin{figure}
     \centering

     \begin{subfigure}[b]{0.4\textwidth}
\includegraphics[width=50mm,scale=0.5,angle=-90]{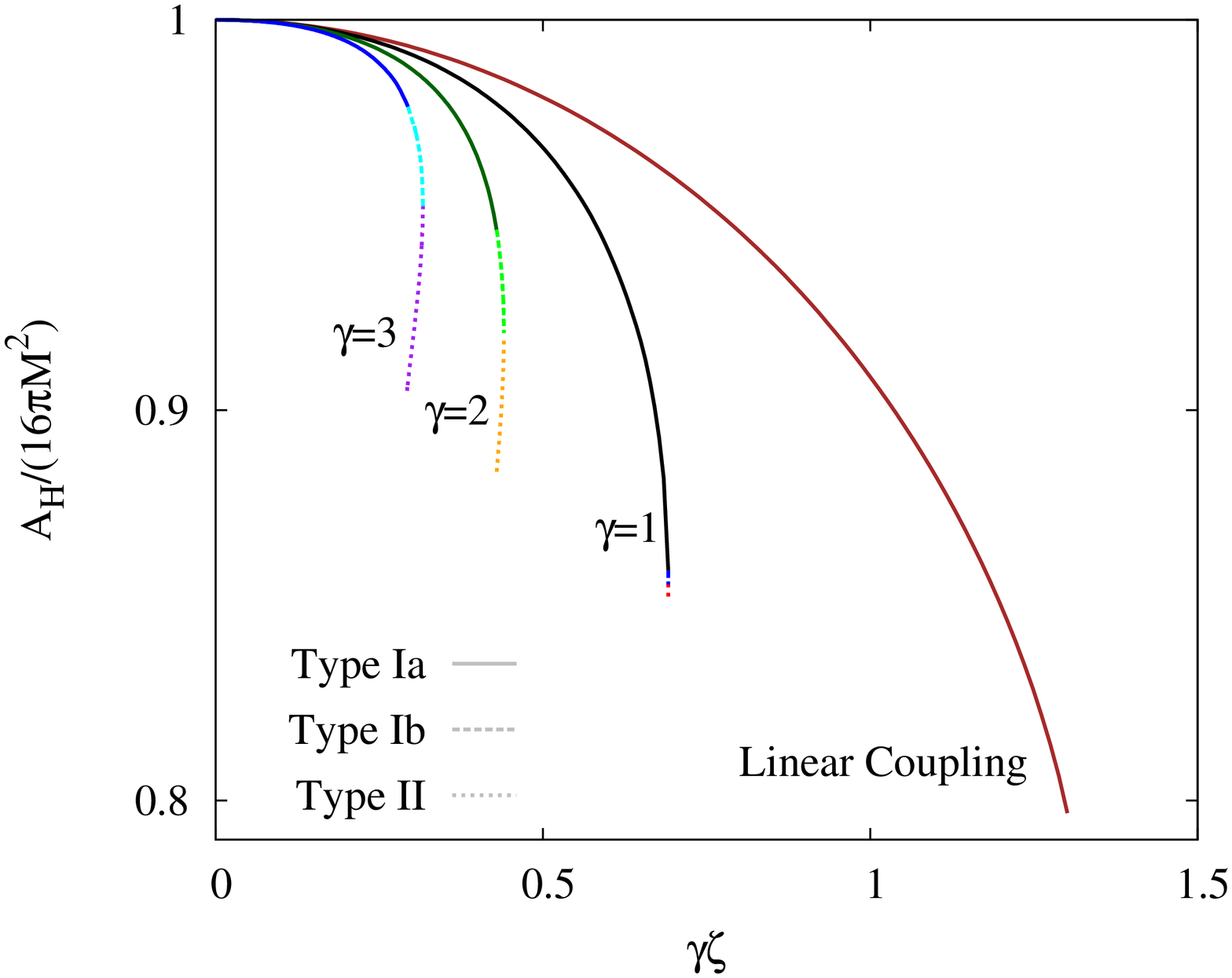}
         \caption{}
         \label{fig:gamma_2branches_a}
     \end{subfigure}
     \begin{subfigure}[b]{0.4\textwidth}
\includegraphics[width=50mm,scale=0.5,angle=-90]{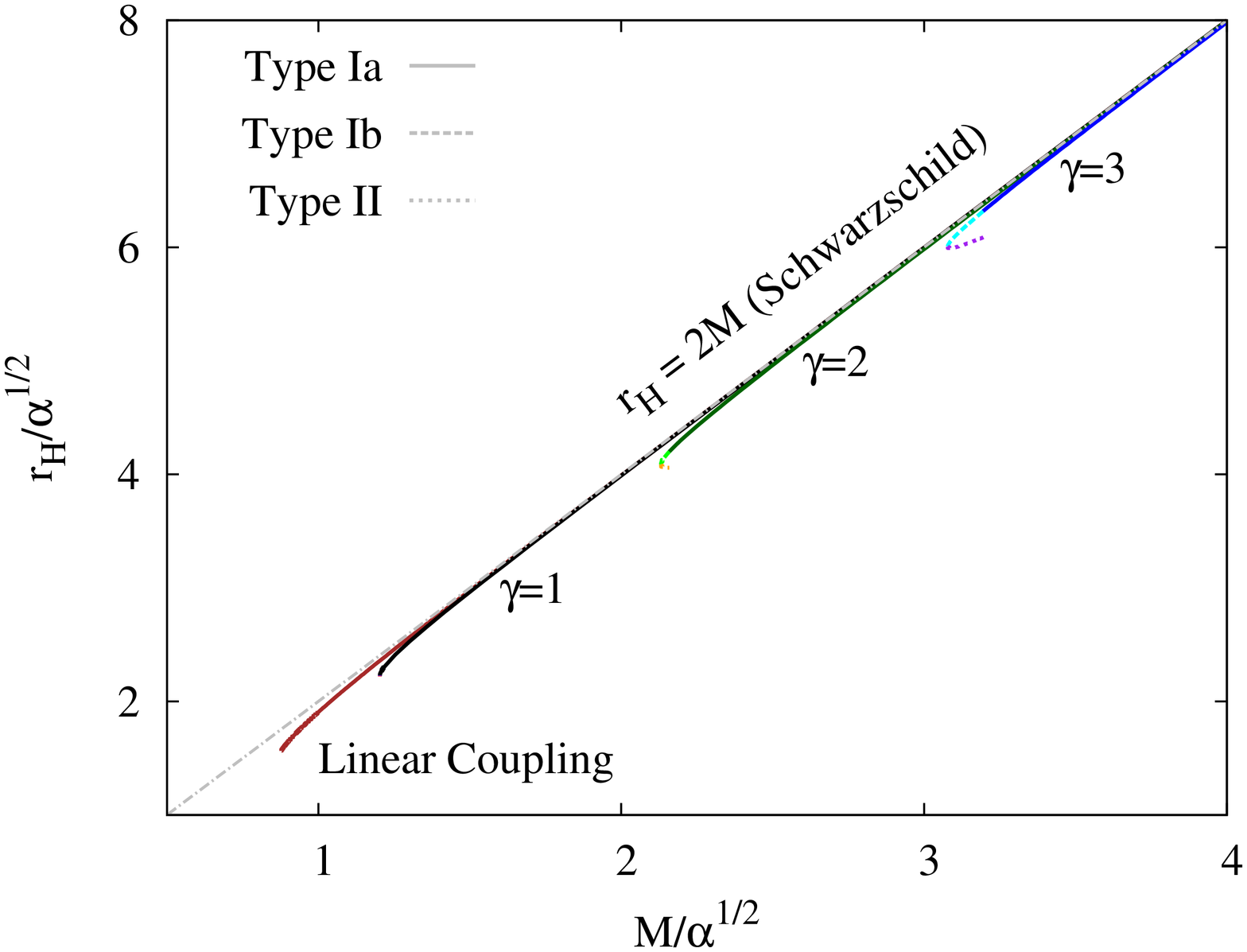}
         \caption{}
         \label{fig:gamma_2branches_b}
     \end{subfigure}

     \caption{(a) The scaled area of the horizon $A_H/16 \pi M^2$
of static EGBd black holes
versus the scaled GB coupling constant $\gamma\zeta$ 
for several values of the dilaton coupling, $\gamma=1, 2, 3$, and the linear coupling case.
(b) The horizon radius $r_H$ versus the mass $M$, both scaled with
respect to the coupling constant $\alpha$, 
for several values of the dilaton coupling, $\gamma=1, 2, 3$, and the linear coupling case.
For comparison the Schwarzschild limit $r_H=2M$ is also shown.}

         \label{fig:gamma_2branches}

\end{figure}

In Fig.\ref{fig:gamma_2branches_a} 
we exhibit the scaled black hole horizon area $A_H/16 \pi M^2$
versus $\gamma\zeta$ for $\gamma=1, 2$ and $3$ and the linear coupling case. 
Again, the type Ia, Ib and II branches are presented 
by different colors and line styles.
In particular, it can be seen that the size of the domain of existence 
decreases with $\gamma$ 
(since $\zeta_L$ decreases as $\gamma$ increases), 
while the relative size of the second branch increases with $\gamma$. 
Note that in the linear coupling case, there are no type II black holes, but the domain of existence is also bounded.

The fact that the domain of existence is bounded by $\zeta_L$ 
means that there is a minimum size for
the static black holes of EGBd theory,
when particular values of $\alpha$ and $\gamma$ are selected.
To show this we present in Fig.\ref{fig:gamma_2branches_b}
the scaled horizon radius $r_H/\sqrt{\alpha}$ 
versus the scaled mass $M/\sqrt{\alpha}$ for three values of $\gamma$ 
and the linear coupling case. 
For a given $\gamma$,
the configuration residing at $\zeta_L$ has the minimal value of
the mass, while the configuration at $\zeta_C$ has again a higher mass.
Note, that for $\gamma=3$ the scaled horizon radius is increasing again
on the secondary branch (for fixed $\alpha$). 

Although in this paper we will focus on these static configurations, 
the effect of rotation on the properties of the EGBd black holes 
has been studied both perturbatively
\cite{PhysRevD.79.084031,Pani:2011gy,Ayzenberg:2014aka,Maselli:2015tta},
and nonperturbatively
\cite{PhysRevLett.106.151104,Kleihaus:2014lba,Kleihaus:2015aje}.
In the rotating case the domain of existence is also bounded,
and the boundaries are given by the Kerr black holes, by the static
EGBd black holes, by the set of critical EGBd black holes
and by the set of extremal EGBd black holes.
With increasing scaled angular momentum $J/M^2$ the 
associated value of $\zeta_L$ decreases, until the
Kerr bound $J/M^2=1$ is reached. Beyond the Kerr bound,
there is also a finite lower limit for $\zeta$, which
merges with $\zeta_L$ at the configuration,
where the maximum value of $J/M^2$ is attained.

\section{Perturbations of Einstein-Gauss-Bonnet-dilaton black holes} \label{sec_QNM}

\subsection{Setup}

In this section we lay out the quasinormal mode formalism, 
and the method used to study the QNM spectrum.

Since there are two dynamical fields in the theory, we perturb the metric and dilaton field separately as 
\begin{eqnarray}
g_{\mu\nu} = g_{\mu\nu}^{(0)}(r) + \epsilon h_{\mu\nu}(t,r,\theta,\varphi)~, \\
\label{Phiperturb}
\Phi = \Phi_{0}(r) + \epsilon \delta \Phi(t,r,\theta,\varphi)~,
\end{eqnarray}
where $\epsilon<<1$ is the perturbation parameter. 
The zeroth order of each of the configurations is a static, spherically symmetric background
described in the previous section, while the perturbations are time-dependent and not spherically symmetric. 
 
Assuming that the nontrivial perturbation functions 
can be decomposed into the product of radial, temporal and angular parts, we 
use bases of spherical harmonic tensors 
to decouple the perturbations into two classes: 
the axial and polar perturbations
\cite{RevModPhys.52.299,ReggeW,PhysRevLett.24.737,FernandezJambrina:2003mv}.
Under a parity transformation, the spherical harmonics
of the axial perturbations transform according to
$
Y_{lm}(\theta,\varphi) \rightarrow Y_{lm}(\pi-\theta,\pi+\varphi) =
(-1)^{l+1}Y_{lm}(\theta,\varphi)
$,
while for polar perturbations they transform as
$
Y_{lm}(\theta,\varphi) \rightarrow Y_{lm}(\pi-\theta,\pi+\varphi) =
(-1)^{l}Y_{lm}(\theta,\varphi)$.
Since these two types of perturbations never mix, 
both channels can be studied separately as in the GR case.

For the axial part of the metric perturbations, 
using a Laplace transformation of the time dependence, 
the expression of the perturbed part of the metric is 
\begin{equation}
h_{\mu\nu}^{(\text{axial})} =\int d\omega \, e^{-i\omega t}  \sum\limits_{l,m}
\left[
\begin{array}{c c c c}
	0 & 0 & -h_{0}	\frac{1}{\sin\theta}\frac{\partial}{\partial\varphi}Y_{lm} & h_{0}	\sin\theta\frac{\partial}{\partial\theta}Y_{lm} \\
	0 & 0 & -h_{1}	\frac{1}{\sin\theta}\frac{\partial}{\partial\varphi}Y_{lm}  
	& h_{1}	\sin\theta\frac{\partial}{\partial\theta}Y_{lm} \\
-h_{0}	\frac{1}{\sin\theta}\frac{\partial}{\partial\varphi}Y_{lm} & 
	-h_{1}	\frac{1}{\sin\theta}\frac{\partial}{\partial\varphi}Y_{lm} 
	& h_2	\frac{1}{2\sin\theta}X_{lm}  & -\frac{1}{2}h_2 \sin\theta \, W_{lm}
 \\
h_{0}	\sin\theta\frac{\partial}{\partial\theta}Y_{lm} & h_{1}	\sin\theta\frac{\partial}{\partial\theta}Y_{lm}  
   & -\frac{1}{2}h_2 \sin\theta \, W_{lm} & -\frac{1}{2} h_2	\sin\theta \, X_{lm}	
\end{array}
\right]
\end{equation}
in the order of $(t,r,\theta,\varphi)$ in the rows and columns of the matrix,
and we have defined the angular functions 
\begin{equation}
W_{lm}
= \frac{\partial^2}{\partial\theta^2}Y_{lm} - \cot\theta \frac{\partial}{\partial\theta}Y_{lm}
-\frac{1}{\sin^2\theta}\frac{\partial^2}{\partial\varphi^2}Y_{lm}~, 
\quad
X_{lm} = 2\frac{\partial^2}{\partial\theta\,\partial\varphi}Y_{lm} - 2 \cot\theta \frac{\partial}{\partial\varphi}Y_{lm}~.
\end{equation}
Note that $h_{0}, h_{1},h_2$ are functions of $r$, 
which carry the integer angular numbers $l,m$, 
and the complex frequency $\omega$. 
The gauge freedom allows us to fix $h_2=0$, 
which is known as the Regge-Wheeler gauge.

On the other hand, for polar perturbations in the metric, we have
\begin{equation}
h_{\mu\nu}^{(\text{polar})} = \int d\omega \, e^{-i\omega t}  \sum\limits_{l,m}
\left[
\begin{array}{c c c c}
	2N F(r) Y_{lm} & -H_1Y_{lm} & 
	-h_{0p}\frac{\partial}{\partial\theta}Y_{lm} & -h_{0p}\frac{\partial}{\partial\varphi} Y_{lm} \\
	-H_1Y_{lm} & -2K(r)LY_{lm} & 
	h_{1p}\frac{\partial}{\partial\theta}Y_{lm} & h_{1p}\frac{\partial}{\partial\varphi}Y_{lm} \\
	-h_{0p}\frac{\partial}{\partial\theta}Y_{lm} & h_{1p} \frac{\partial}{\partial\theta}Y_{lm} 
	& B & -r^2VX_{lm}
	\\
	 -h_{0p} \frac{\partial}{\partial\varphi} Y_{lm} & h_{1p}\frac{\partial}{\partial\varphi} Y_{lm}  & 
  -r^2 V X_{lm} & A \\
\end{array}
\right]~,
\end{equation}
where 
$A = (l(l+1)V - 2 T)r^2 \sin^2\theta \, Y_{lm} + r^2 V \sin^2\theta\,W_{lm}$ and 
$B = (l(l+1)V - 2 T)r^2  Y_{lm} - r^2 V W_{lm}$. 
Recall that $F(r), K(r)$ are the metric functions from Eq.(\ref{ds2}).
The functions $N, V, T, L, H_1, h_{0p}, h_{1p}$ depend on the radial coordinate 
$r$, the angular numbers $l$, $m$, and the complex frequency $\omega$. 
Similarly, we can choose the gauge-fixing $h_{0p}=h_{1p}=V=0$.

Since the dilaton is a scalar field, it only experiences polar perturbations.
Factorizing the perturbation function in a similar way we have
\begin{equation} 
\delta \Phi =  \int d\omega \, e^{-i\omega t} \sum_{l,m} \Phi_1 \, Y_{lm}~,
\end{equation}
where $\Phi_1$ depends on $r,l,m$ and $\omega$, like the metric
perturbation functions.

Note that in our convention, the temporal part of the perturbations 
is taken to be $e^{-i\omega t}$, where the wave frequency $\omega$ is 
a complex number, 
$\omega = \omega_{R} + i\omega_{I}$ for $\omega_R, \omega_I \in \mathbb{R}$.
By using this ansatz for the axial and polar perturbations, 
we are able to decouple the differential equations into a set of 
ordinary differential equations in $r$ with an undetermined frequency $\omega$.

Before proceeding to the analysis of the field equations,
we should specify the proper behaviour of the waves 
close to the horizon and to infinity.
In the tortoise coordinate $r_{*}$, with $dr_{*}/dr= \sqrt{K(r)/F(r)}$,
the waves should be purely ingoing at the horizon 
and purely outgoing at infinity.
Approaching the horizon ($r_{*}\rightarrow - \infty$),
the radial part of the waves goes as $e^{-i\omega r_{*}}$ 
and toward infinity ($r_{*} \rightarrow \infty$) it goes as
as $e^{i\omega r_{*}}$.
Considering the time dependence 
$e^{-i\omega t} = e^{-i\omega_{R} t} e^{\omega_{I}t}$, 
the imaginary part $\omega_I$ corresponds to $1/\tau$,
where $\tau$ is the damping time.
The waves will exponentially decay in time, when $\omega_I$ is negative.

\subsection{Equations, boundary conditions and numerical method}

The modified Einstein equations up to first order perturbations can be reduced to a minimal set of differential equations of the form:

\begin{equation}
\frac{d}{dr}{{\Psi}}_{(i)}+{U}_{(i)}{\Psi}_{(i)}=0~,
\label{eq_perturbations}
\end{equation}
where $(i)=axial$, $polar$. In the axial case $\Psi_{axial} = (h_0, h_1)$ and in the polar case $\Psi_{polar} = (H_1, T, \Phi_1, \frac{d}{dr}\Phi_1)$. The matrix $U_{(i)}$ contains the coefficients of the equations, which are given by combinations of the functions of the static solution $F(r), K(r)$, the $l$ number
(there is degeneracy with respect to $m$), 
and the frequency eigenvalue $\omega$. 
Since these equations are lengthy, we refer the readers to previous papers,
where the expressions have already been presented 
\cite{PhysRevD.79.084031,Blazquez-Salcedo:2016enn,Blazquez-Salcedo:2015ets}.

\begin{figure}
     \centering

\includegraphics[width=50mm,scale=0.5,angle=-90]{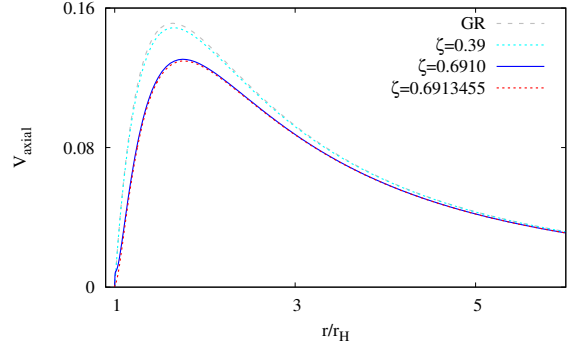}

     \caption{The potential $V_{axial}$ ($l=2$) as a function of the radial coordinate $r$, for $\gamma=1$ and different values of the coupling $\zeta$ including the GR case.}

         \label{fig:Vaxial}

\end{figure}

{In principle, 
Eq.(\ref{eq_perturbations}) can be packaged into a single time-independent 
second order differential equation, which takes the form of the 
Regge-Wheeler/Zerilli equation,
$d^2 Z_{(i)}/dr^2_{*} + (\omega^2 - V_{(i)}) Z_{(i)} = 0$, 
with $V_{(i)}$ the potential. 
In our case, $Z_{axial}$ is a combination of functions $h_0, h_1$, 
which describes the axial perturbations.
For the polar perturbations we have two coupled functions in 
$Z_{polar}=(Z,\Phi_1)$, where $Z$ is a combination of the functions $H_1, T$.
As an example, we show in Figure \ref{fig:Vaxial} the potential $V_{axial}$ as a function of the radial coordinate $r$, for different values of the $\zeta$ parameter.
Nonetheless, the first order differential equation system 
from Eq.(\ref{eq_perturbations}) is sufficient for our purpose here.
}

The outgoing wave behavior of the perturbations at infinity imposes the following behavior on the axial perturbations:
\begin{eqnarray}
h_0(r) & \approx & e^{i\omega r_*} \cdot [h_{00} + \frac{h_{01}}{r} + O(r^{-2})]~, \\
h_1(r) & \approx & e^{i\omega r_*} \cdot [-h_{00} + \frac{i}{\omega}((2iM\omega-1)h_{01}+i\omega h_{00})\frac{1}{r} + O(r^{-2})]~,
\end{eqnarray}
where the parameter $h_{01}$ satisfies
\begin{equation}
h_{01} = \frac{i h_{00}}{8\omega}(\omega^2(4 f_2+Q^2)+4(l+2)(l-1))~,
\end{equation}
and $f_2$ and $Q$ are determined by the asymptotic expansion of the static metric (see Sec.\ref{sec_theory}).

In the case of polar perturbations, the functions behave like:
\begin{eqnarray}
T(r) &\approx & e^{i\omega r_*} \cdot [T_0 + O(r^{-2})]~, \\
H_1(r) &\approx & \omega r e^{i\omega r_*} \cdot [-T_{0} -\frac{1}{2\omega r}(4M\omega + i(l^2+l-2))T_{0} + O(r^{-2})]~, \\
\Phi_1(r) &\approx & e^{i\omega r_*} \cdot \frac{1}{r} \cdot [\Phi_{10} - \frac{i}{2\omega r}(2iM\omega Q T_0 - l(l+1)\Phi_{10}) + O(r^{-2})]~.
\end{eqnarray}

At the horizon, we require waves falling into the horizon. This means the axial functions have the following near-horizon expansion:
\begin{eqnarray}
h_0(r) &\approx &\frac{e^{-i\omega r_*}}{r-r_H} \cdot [\hat{h}_{00} + O(r-r_H)]~, \\
h_1(r) &\approx &-\frac{e^{-i\omega r_*}}{r_H} \cdot [\hat{h}_{10} + O(r-r_H)]~.
\end{eqnarray}
The parameter $\hat{h}_{10}$ satisfies:
\begin{eqnarray}
\hat{h}_{10} = -\hat{h}_{00} \sqrt{\frac{\alpha\gamma \Phi_{01} + 2e^{\gamma\Phi_{00}} r_H}{2r_H^2 f_1 e^{\gamma\Phi_{00}}}}~,
\end{eqnarray}
where $\Phi_{00} :=\Phi_0(r_H)$, and the parameters $f_1$ and $\Phi_{01}$ are given by the expansion of the static solution near the horizon.

In the polar case, the behavior close to the horizon is:
\begin{eqnarray}
T(r) &\approx &e^{-i\omega r_*} \cdot [\hat{T}_0 + O(r-r_H)]~, \\
H_1(r) &\approx &e^{-i\omega r_*} \cdot \frac{\omega}{r-r_H} \cdot [\hat{H}_{10} + O(r-r_H)]~, \\
\Phi_1(r) &\approx & e^{-i\omega r_*} 
\cdot [\hat{\Phi}_{10} + O(r-r_H)]~.
\end{eqnarray}
The parameter $\hat{H}_{10}$ satisfies:
\begin{equation}
\hat{H}_{10} = \frac{D_2}{D_1} (e^{-\gamma\Phi_{00}} \alpha \gamma \hat{\Phi}_{10} - r_H^3 \hat{T}_0)~,
\end{equation}
where we define
\begin{eqnarray}
D_2 &:=& (4 e^{\gamma\Phi_{00}} r_H + 2 \alpha \gamma \Phi_{01})\omega^2 + e^{\gamma\Phi_{00}} f_1 ~, \nonumber  \\
D_1 &:=& 2 r_H [\alpha \gamma \Phi_{01} 
+ 2 e^{\gamma\Phi_{00}} r_H]\omega^2 - r_H f_1 e^{\gamma\Phi_{00}} l (l+1) + \nonumber \\ 
&&i \omega r_H e^{\gamma\Phi_{00}} (l^2 + l +1) \sqrt{2 \alpha \gamma e^{-\gamma \Phi_{00}} \Phi_{01} f_1 + 4 r_H f_1}~. 
\end{eqnarray}

In order to generate the quasinormal modes, we use a numerical procedure that implements these differential equations together with boundary conditions that satisfy the previous expansions. The method essentially consists of the following steps:
\\
First we generate numerically a static dilatonic black hole by solving the EGBd equations with the appropriate boundary conditions 
\cite{Kanti:1995vq}.
The estimated relative error of the solutions is lower than $10^{-12}$ for a mesh with more than $1000-2000$ points.
The next step is to choose a value of $l$ and a value of $\omega$. By moving outside the horizon $r_{s1} = r_H(1+\epsilon)$, we evaluate the previous boundary conditions at $r = r_{s1}$. Using a shooting method we generate a first solution of the perturbation equations from $r \in [r_{s1}, r_0]$. Typically, $\epsilon \approx 10^{-5}$. To calculate the coefficients of $U_{(i)}$ of Eq.(\ref{eq_perturbations}), we use the numerical static solution with an interpolation. Similarly, we choose a large value of $r_{s2} >> r_H$, and with a shooting method we generate a second solution from $r \in [r_{0}, r_{s2}]$. Typically $r_{s2} \approx 100 r_H$. Finally, we study the continuity between the first and the second solution at $r_0$, which is typically around $(4-6) r_H$. 
If the solution is continuous within a required precision, 
then we take $\omega$ as the frequency of a quasinormal mode, 
with number $l$, of the black hole under consideration. 
To find such solutions,
we implement a numerical procedure to explore the complex plane 
for $\omega$.
Typically we require $\omega$ to have a relative precision below $10^{-3}$.

In the following sections we present the numerical results,
where we have varied the GB coupling constant $\alpha$,
the dilaton coupling $\gamma$,
the black hole mass $M$, and the angular number $l$.

\section{Results and Discussion} \label{sec_results}

\subsection{Spectrum for fundamental $l=0, 1, 2, 3$ modes 
of the type I black holes}

In this section we will focus on the EGBd black holes
on the primary branch, i.e., type I black holes, 
with a fixed dilaton coupling $\gamma=1$. 
Later (in Sec.\ref{sec_gamma}) 
we will discuss the effect of changing $\gamma$.
We now begin by showing the complete set of results 
for the fundamental $l = 0,..,3$ modes 
of the axial and polar perturbations. 

In GR, gravitational waves are obtained starting from 
$l=2$ (quadrupole radiation), 
where typically $l=2$ and $l=3$ are expected to dominate 
any physically relevant signal.
In addition, one can consider scalar $l=s=0$ modes and 
vector $l=s=1$ modes, 
by studying a field with spin $s$ propagating in the background
of a Schwarzschild or Kerr black hole 
(see e.g.~\cite{Berti:2009kk,Konoplya:2011qq}).

In EGBd theory, the situation changes somewhat, since with the dilaton
field there is already a scalar field present in the background
black hole solution, which can give rise to the
emission of gravitational waves. Thus $l=0$ and $l=1$ modes
are inherent to the theory, and their study is essential
for a complete physical understanding.
In particular, their study is necessary for an analysis
of the linear mode stability.
Considering only the metric and the dilaton field
(and no further external fields),
then all axial perturbations
are gravitationally induced perturbations.
In contrast, for the polar perturbations 
a new channel of gravitational radiation is obtained, 
with modes exciting both spacetime and scalar fluctuations,
since the dilaton is dynamically coupled to the spacetime
\cite{Yagi:2012gp,Blazquez-Salcedo:2016enn}.

\subsubsection{$l=0$ monopole and $l=1$ dipole perturbations}

In the following, we will present the results for the quasinormal modes
in dimensionless quantities,
by exhibiting the scaled frequency $\omega_{R} M$
versus the scaled GB coupling constant $\zeta = \alpha/M^2$.

\begin{figure}
     \centering

     \begin{subfigure}[b]{0.4\textwidth}
\includegraphics[width=50mm,scale=0.5,angle=-90]{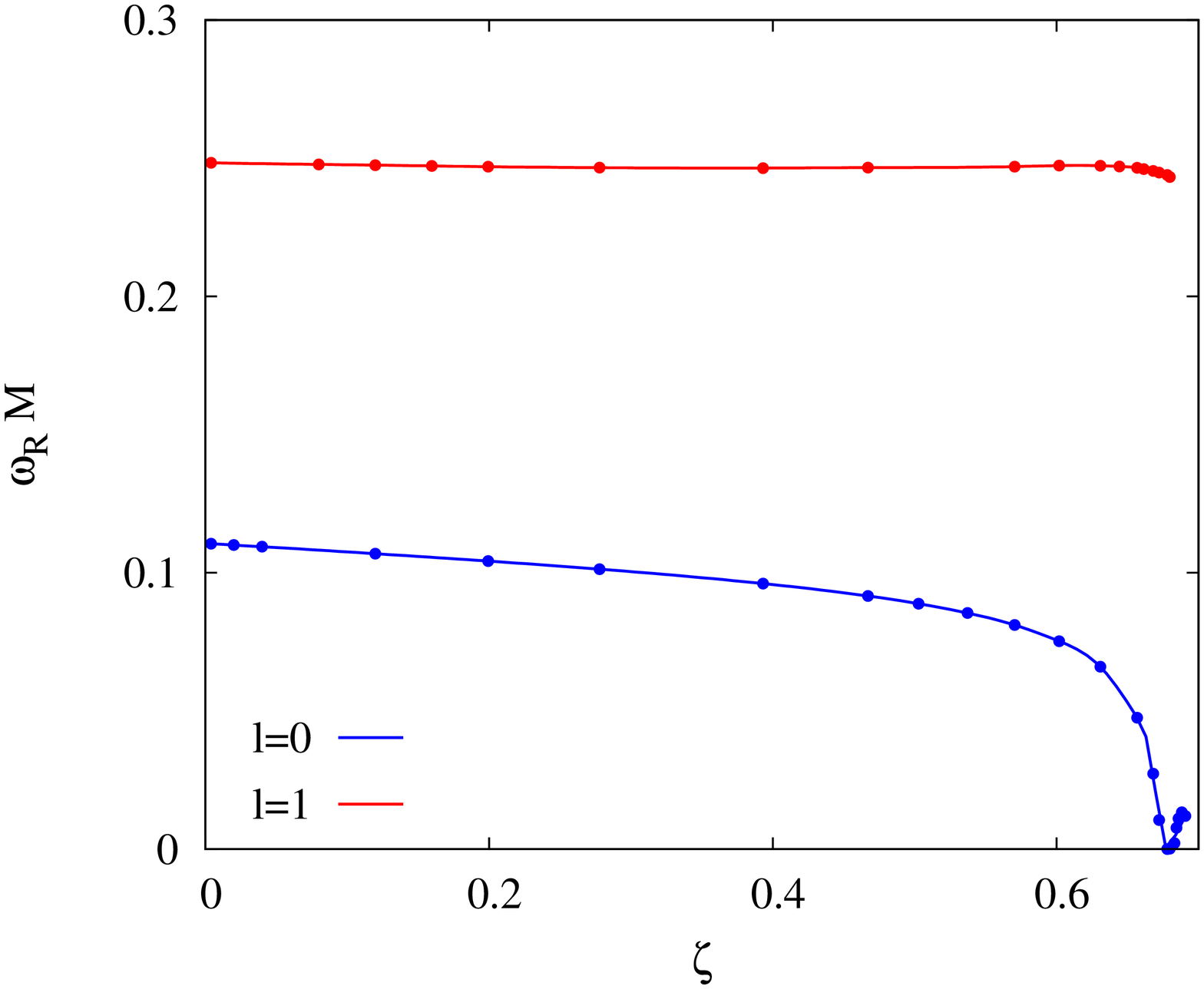}
         \caption{}
         \label{fig:l0l1_a}
     \end{subfigure}
     \begin{subfigure}[b]{0.4\textwidth}
\includegraphics[width=50mm,scale=0.5,angle=-90]{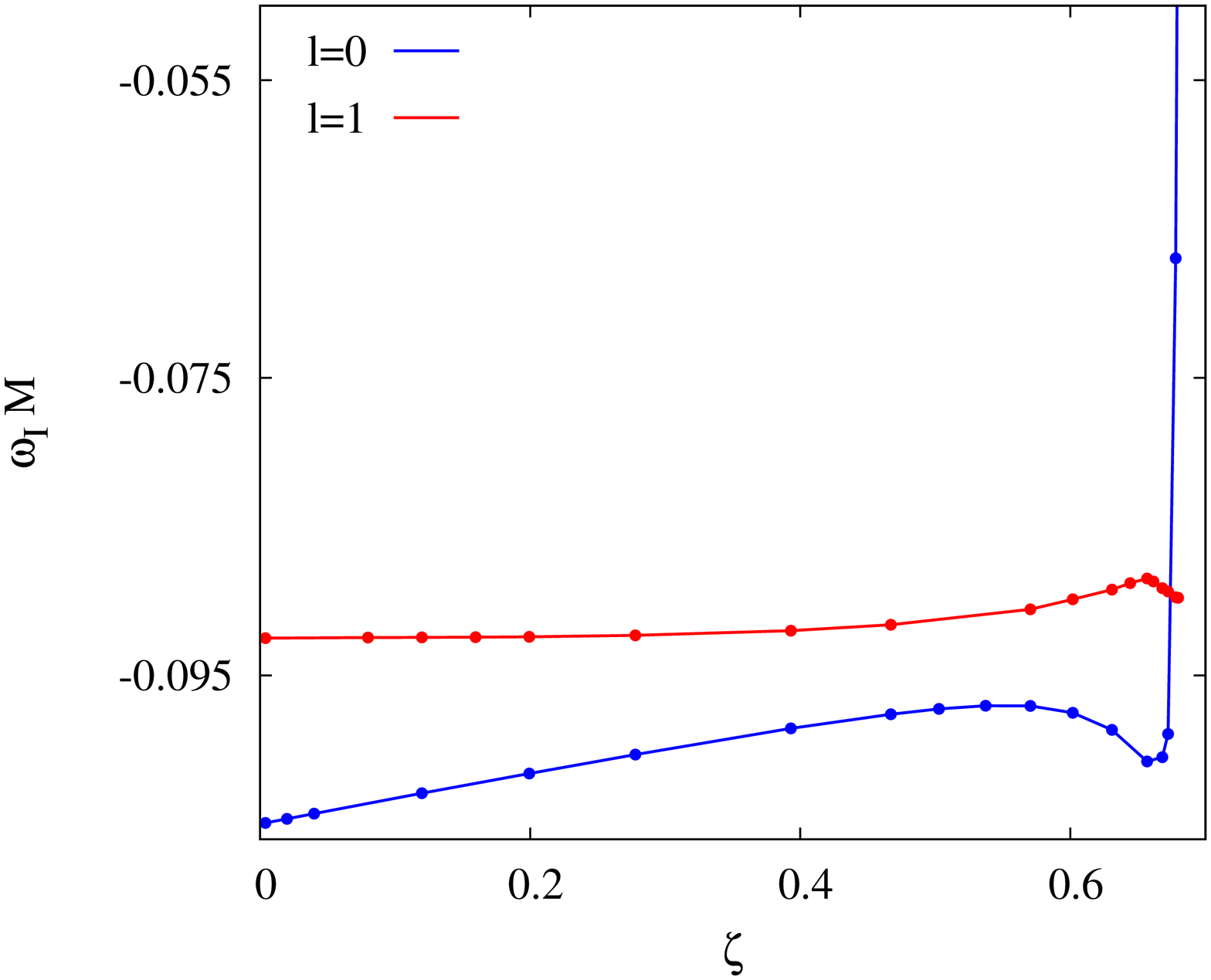}
         \caption{}
         \label{fig:l0l1_b}
     \end{subfigure}

     \caption{(a) Real part of the scaled frequency $\omega_{R} M$ 
as a function of the scaled GB coupling constant $\zeta = \alpha/M^2$
for the fundamental modes with $l=0$ (blue) and $l=1$ (red). 
(b) Imaginary part $\omega_{I} M$ versus $\zeta$ for these modes.}

         \label{fig:l0l1}

\end{figure}

When $\zeta=0$, we start from the corresponding modes
$\omega M = 0.1105 - 0.1049i$ for $l=s=0$ and
$\omega M = 0.2483 - 0.09249i$ for $l=s=1$
of external fields with spin $s$ in the background Schwarzschild metric.
But as soon as $\zeta$ assumes a finite value, 
and the black holes represent genuine EGBd black holes,
the dilaton and metric perturbations are coupled.
Thus the waves emitted in these channels
carry a combination of dilaton and metric excitations.

In Fig.\ref{fig:l0l1} we present the frequencies of
the fundamental $l=0$ and $l=1$ EGBd modes, 
where the real part $\omega_{R} M$
is shown in Fig.\ref{fig:l0l1_a} and the
imaginary part $\omega_{I} M$ is shown in Fig.\ref{fig:l0l1_b}.

For a fixed mass, the real part $\omega_{R}$ of the $l=0$ mode
first decreases slightly from the Schwarzschild value
with increasing $\zeta$, until
around $\zeta \approx 0.6$, 
$\omega_{R}$ shows a sharp decline, 
approaching zero at $\zeta = 0.678$.
However, $\omega_{R}$ rises again beyond $\zeta = 0.678$.
(Recall the maximal value $\zeta_L = 0.69137$.)
In particular, black holes close to the 
limiting coupling constant $\zeta_L$ have a nonvanishing 
real part of the frequency, $\omega_{R}$,
as we will show in Sec.\ref{2nd-branch}. 
The imaginary part of the frequency $\omega_{I}$
first increases slightly from the Schwarzschild value,
then reaches a local maximum and next a local minimum,
and then rises sharply, as it gets closer toward $\zeta_L$.
It is important to note that for type I black holes, 
the value of $\omega_{I}$ remains always negative,
which is important for their stability.

For $l=1$ the $\zeta$-dependence is very different. 
The real part of the frequency $\omega_{R}$ 
first decreases very slowly from the Schwarzschild value.
For intermediate values of the coupling constant $\zeta$
it slightly rises again, and then shows another
small decrease toward $\zeta_L$.
The same applies to the imaginary part of the frequency $\omega_{I}$. 
Overall the $l=1$ mode is only very weakly dependent on $\zeta$. 
Thus the deviations from the $l=1$ mode in GR
are rather small.

\subsubsection{Axial fundamental modes for $l=2$ and $l=3$}

\begin{figure}
     \centering

     \begin{subfigure}[b]{0.4\textwidth}
\includegraphics[width=50mm,scale=0.5,angle=-90]{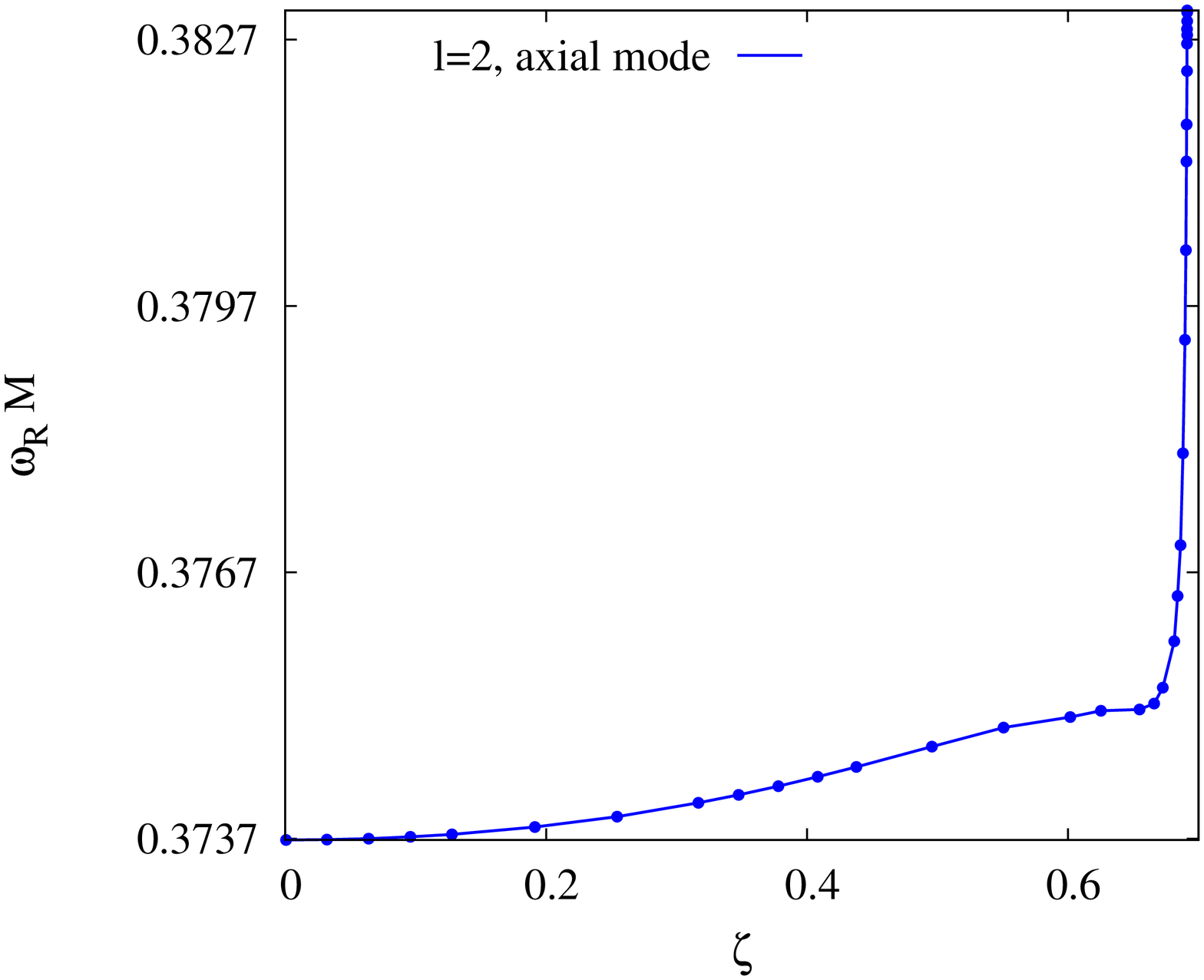}
         \caption{}
         \label{fig:l2_axial_a}
     \end{subfigure}
     \begin{subfigure}[b]{0.4\textwidth}
\includegraphics[width=50mm,scale=0.5,angle=-90]{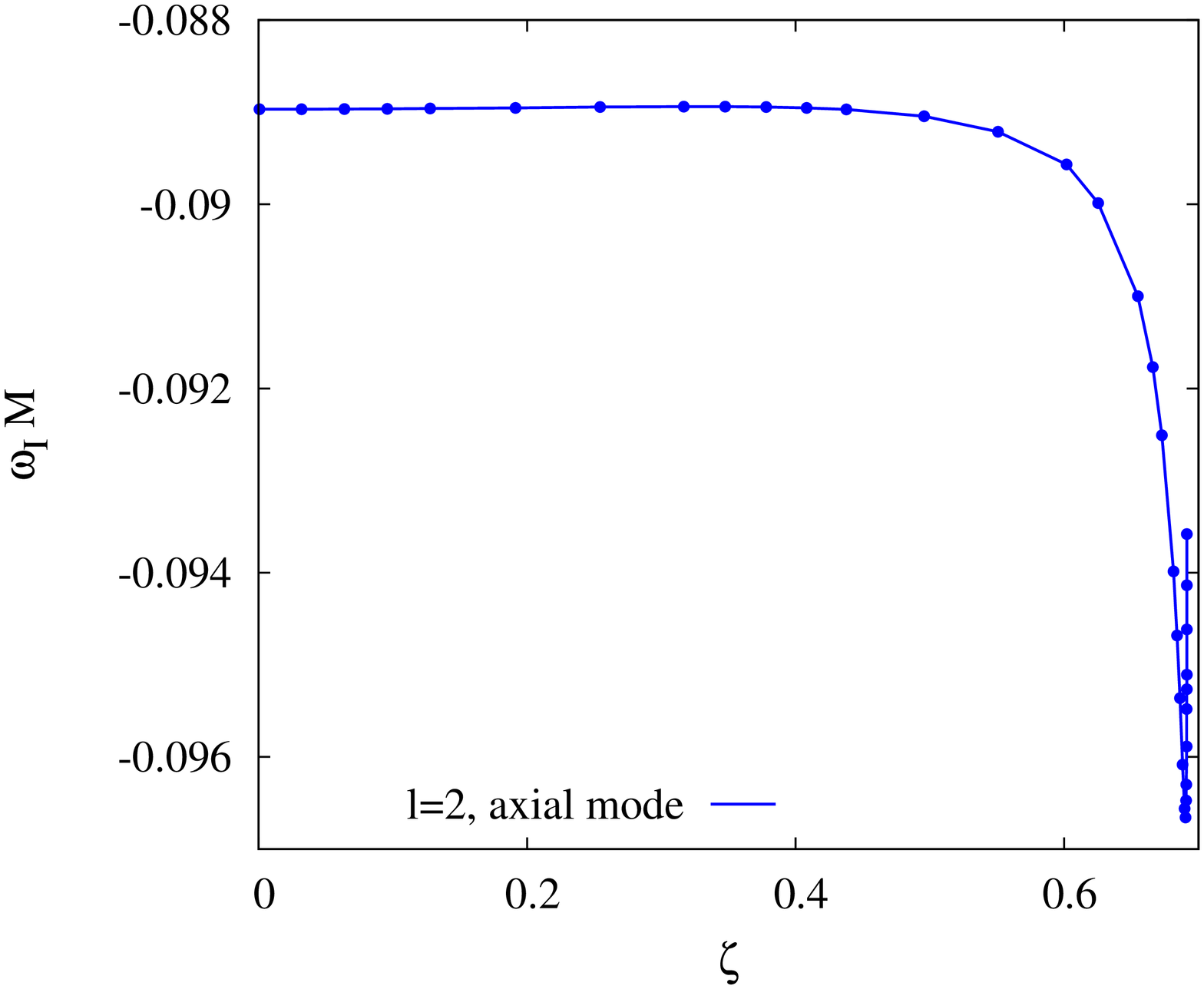}
         \caption{}
         \label{fig:l2_axial_b}
     \end{subfigure}

     \begin{subfigure}[b]{0.4\textwidth}
\includegraphics[width=50mm,scale=0.5,angle=-90]{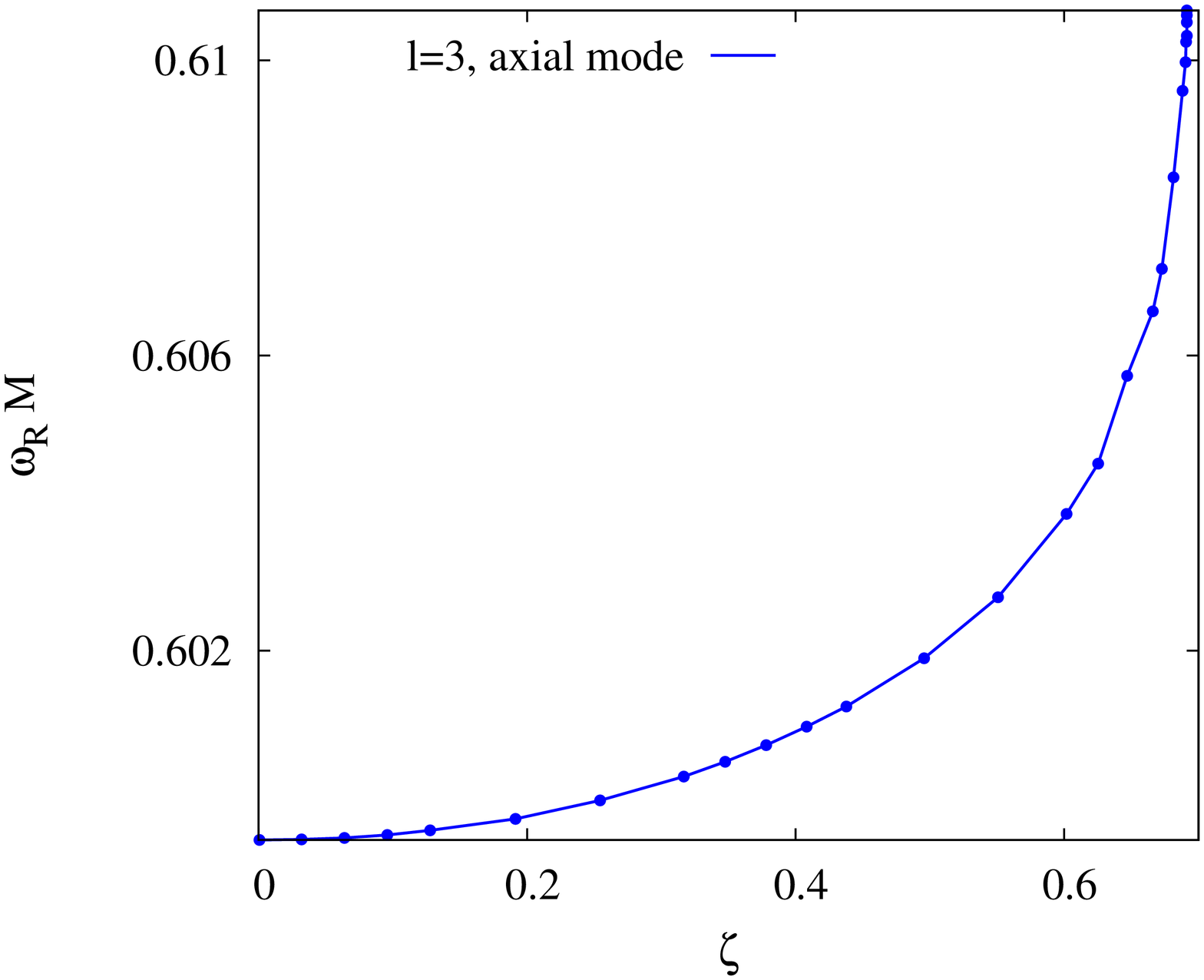}
         \caption{}
         \label{fig:l3_axial_a}
     \end{subfigure}
     \begin{subfigure}[b]{0.4\textwidth}
\includegraphics[width=50mm,scale=0.5,angle=-90]{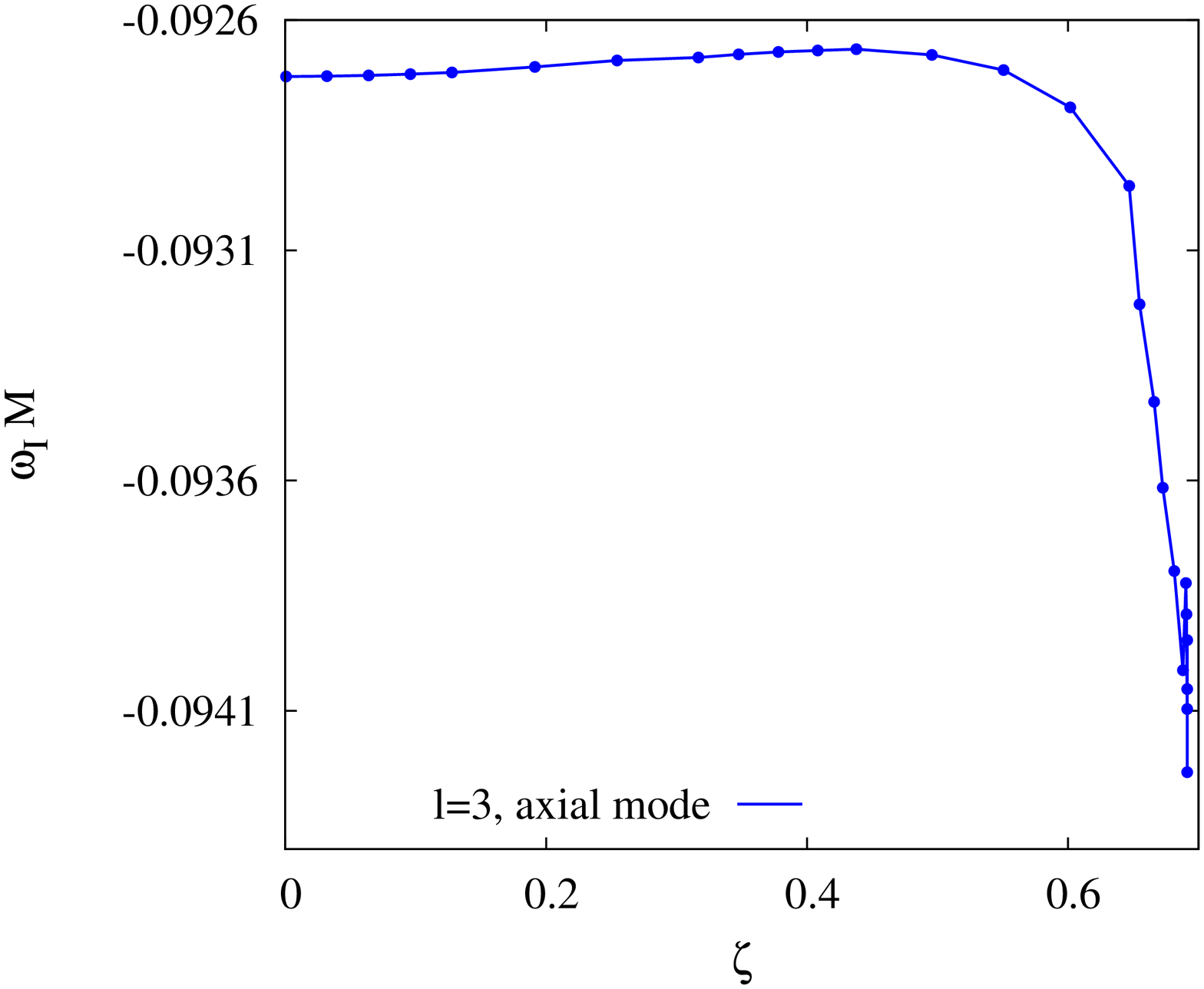}
         \caption{}
         \label{fig:l3_axial_b}
     \end{subfigure}

     \caption{Scaled frequency $\omega M$
of the axial fundamental modes with $l=2$ and $l=3$
as a function of the scaled GB coupling constant $\zeta = \alpha/M^2$.
(a) Real part $\omega_R M$ for $l=2$, 
(b) imaginary part $\omega_I M$ for $l=2$, 
(c) real part $\omega_R M$ for $l=3$, 
(d) imaginary part $\omega_I M$ for $l=3$.} 

         \label{fig:l2l3_axial}

\end{figure}

Let us now present the axial fundamental $l=2$ and $l=3$ modes. 
Axial perturbations describe pure spacetime perturbations.
Therefore these perturbations are purely gravitational in nature 
with no excitation of the dilaton field.
In principle such modes could be excited in a merger process.

Let us first recall the known GR values of the frequency 
for the fundamental modes. These are
$\omega M = 0.3737 - 0.08895i$ for $l=2$, 
and $\omega M = 0.5994 - 0.09270i$ for $l=3$.
To convert these values into Hz, 
one has to multiply by 
$2\pi \times 5142 \,\text{Hz} \times M_{\odot}/M$, 
where $M_{\odot}$ is the solar mass \cite{Kokkotas1999}.

We present the frequencies of the axial fundamental $l=2,\ 3$ 
EGBd modes in Fig.\ref{fig:l2l3_axial} 
versus the scaled coupling constant $\zeta$. 
The left figures show the real part of the frequency, 
while the imaginary part is shown on the right.

Fig.\ref{fig:l2_axial_a} shows
that the real part $\omega_R M$ increases 
slightly with the coupling constant for $l=2$. 
This holds everywhere
except in a small region close to $\zeta_L$, 
which we will explore in detail in Sec.\ref{2nd-branch}. 
The deviation from GR is very small, 
less than $0.5\%$ below $\zeta = 0.65$, 
and only up to $2.5 \%$ beyond.
The imaginary part $\omega_{I} M$ is shown in 
Fig.\ref{fig:l2_axial_b}.
Again, the deviation from GR is very small 
for smaller values of the coupling.
Around $\zeta=0.6$ the mode becomes damped faster 
(the difference to GR rising up to $8\%$). 
Close to $\zeta_L$ the dependence of $\omega_{I} M$ 
becomes more complicated,
and we will zoom into this region in Sec.\ref{2nd-branch}.

In Fig.\ref{fig:l3_axial_a} and \ref{fig:l3_axial_b}, 
we show the analogous figures for the axial fundamental $l=3$ mode 
as a function of $\zeta$. 
In Fig.\ref{fig:l3_axial_a} we see that the real part $\omega_R M$ 
increases quadratically with $\zeta$ up to $1.5\%$ toward $\zeta_L$. 
The imaginary part, shown in Fig.\ref{fig:l3_axial_b}, however, 
hardly deviates from the GR value for intermediate values of $\zeta$, 
and it only decreases down to $2\%$, when $\zeta>0.6$ 
in the vicinity of $\zeta_L$. 
Overall, the variation of the axial fundamental $l=3$ mode 
with the coupling constant is quite small as compared to the $l=2$ mode.

\subsubsection{Polar fundamental modes for $l=2$ and $l=3$}

\begin{figure}
     \centering

     \begin{subfigure}[b]{0.4\textwidth}
\includegraphics[width=50mm,scale=0.5,angle=-90]{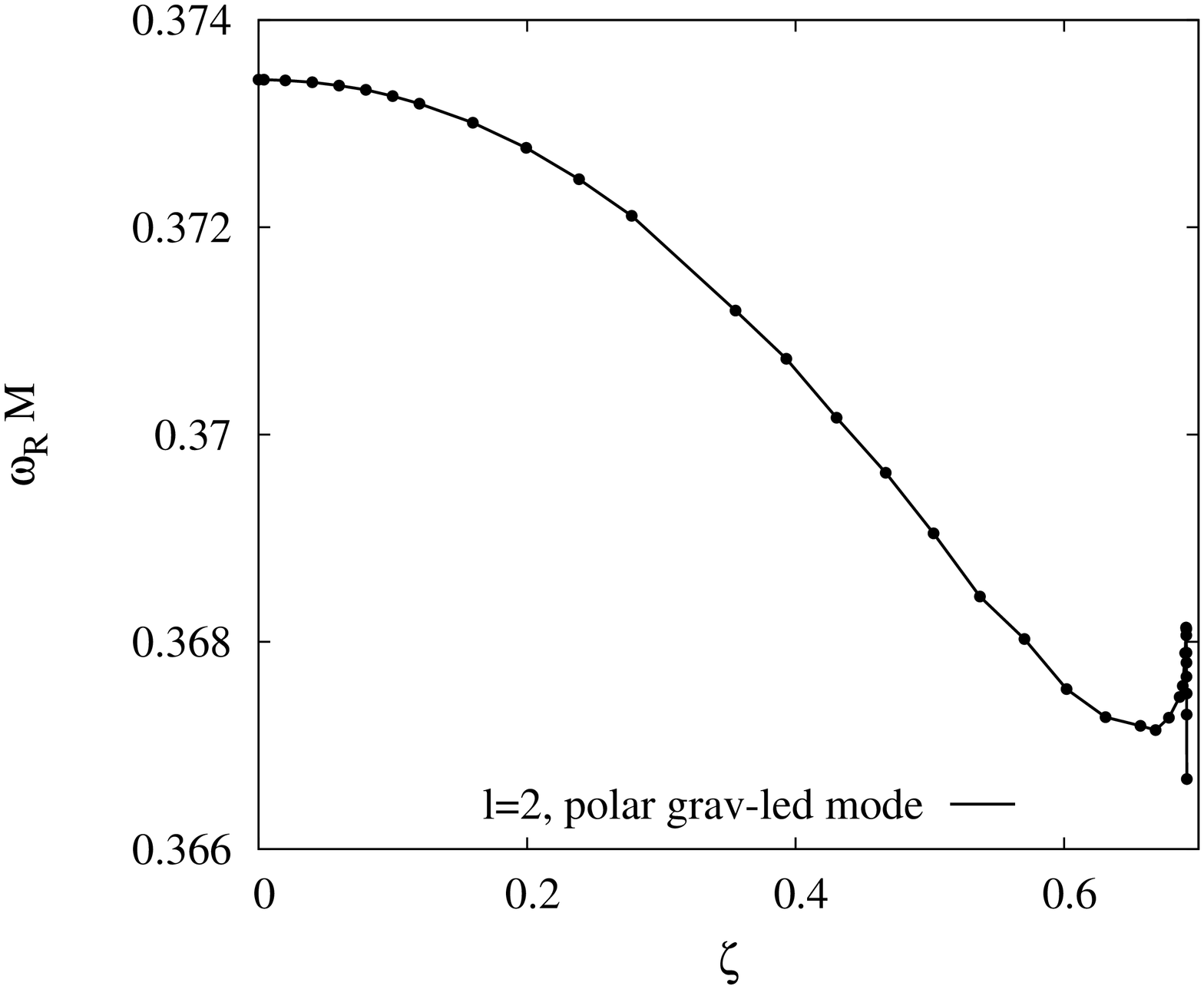}
         \caption{}
         \label{fig:gravitational_a}
     \end{subfigure}
     \begin{subfigure}[b]{0.4\textwidth}
\includegraphics[width=50mm,scale=0.5,angle=-90]{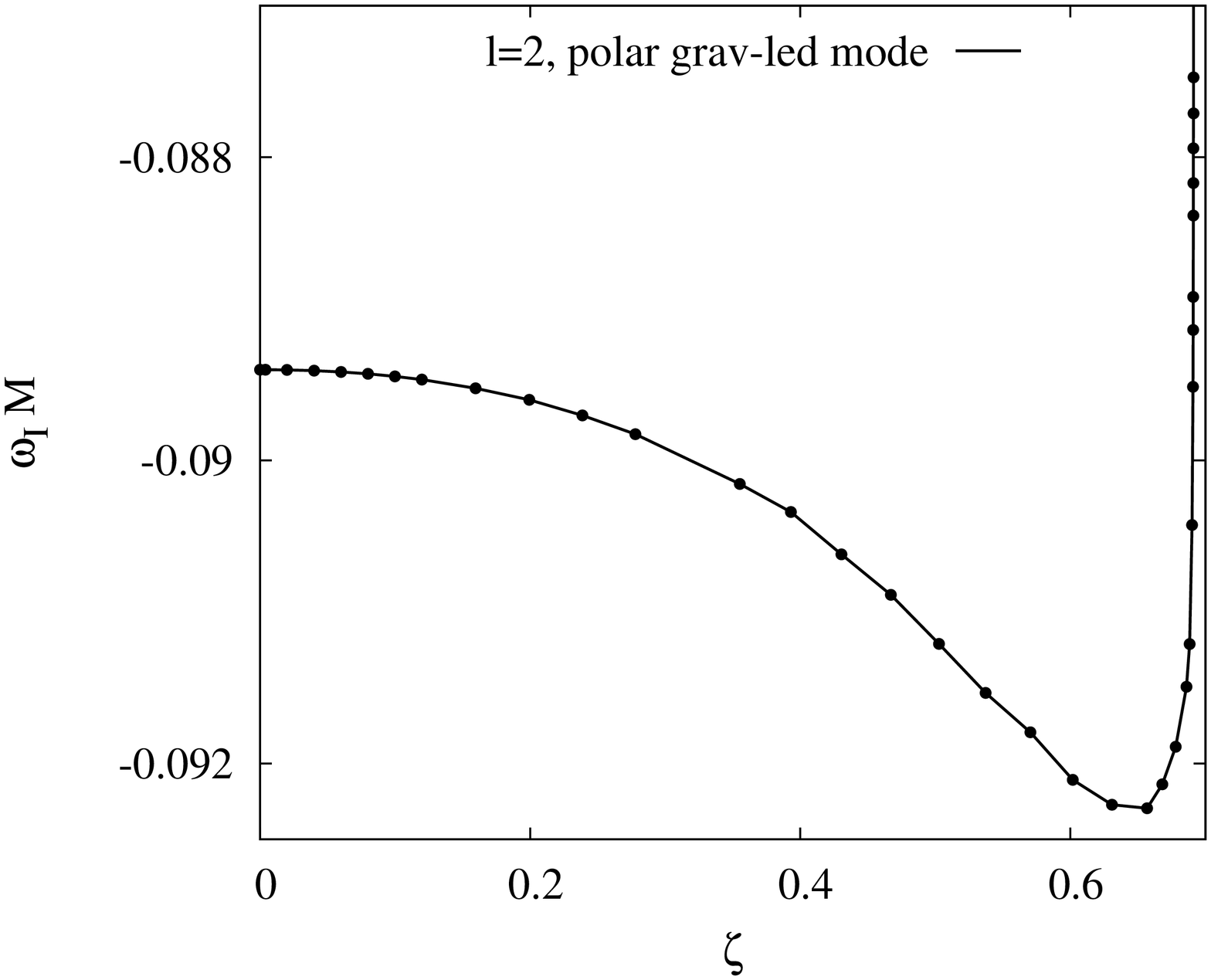}
         \caption{}
         \label{fig:gravitational_b}
     \end{subfigure}

     \begin{subfigure}[b]{0.4\textwidth}
\includegraphics[width=50mm,scale=0.5,angle=-90]{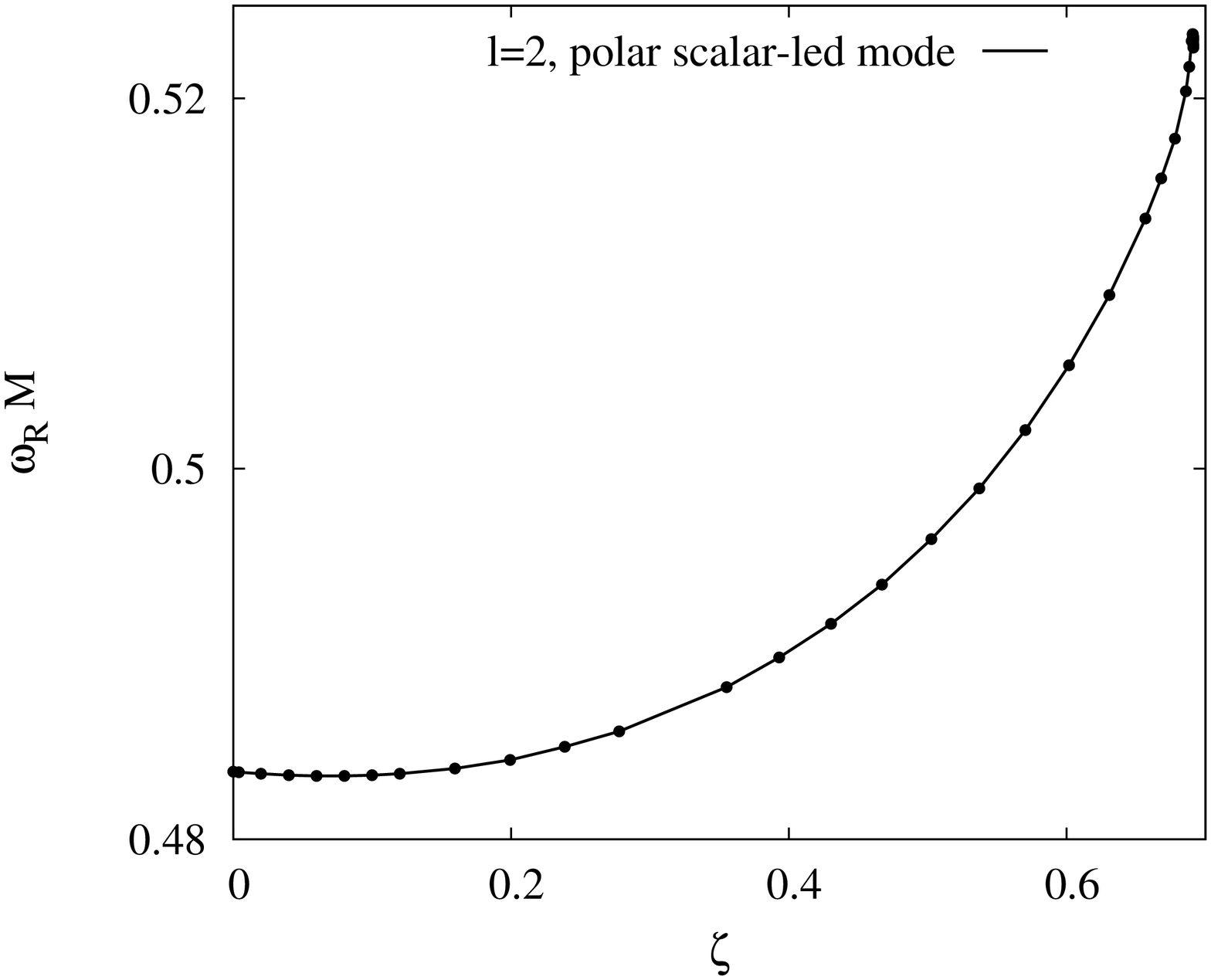}
         \caption{}
         \label{fig:dilaton_a}
     \end{subfigure}
     \begin{subfigure}[b]{0.4\textwidth}
\includegraphics[width=50mm,scale=0.5,angle=-90]{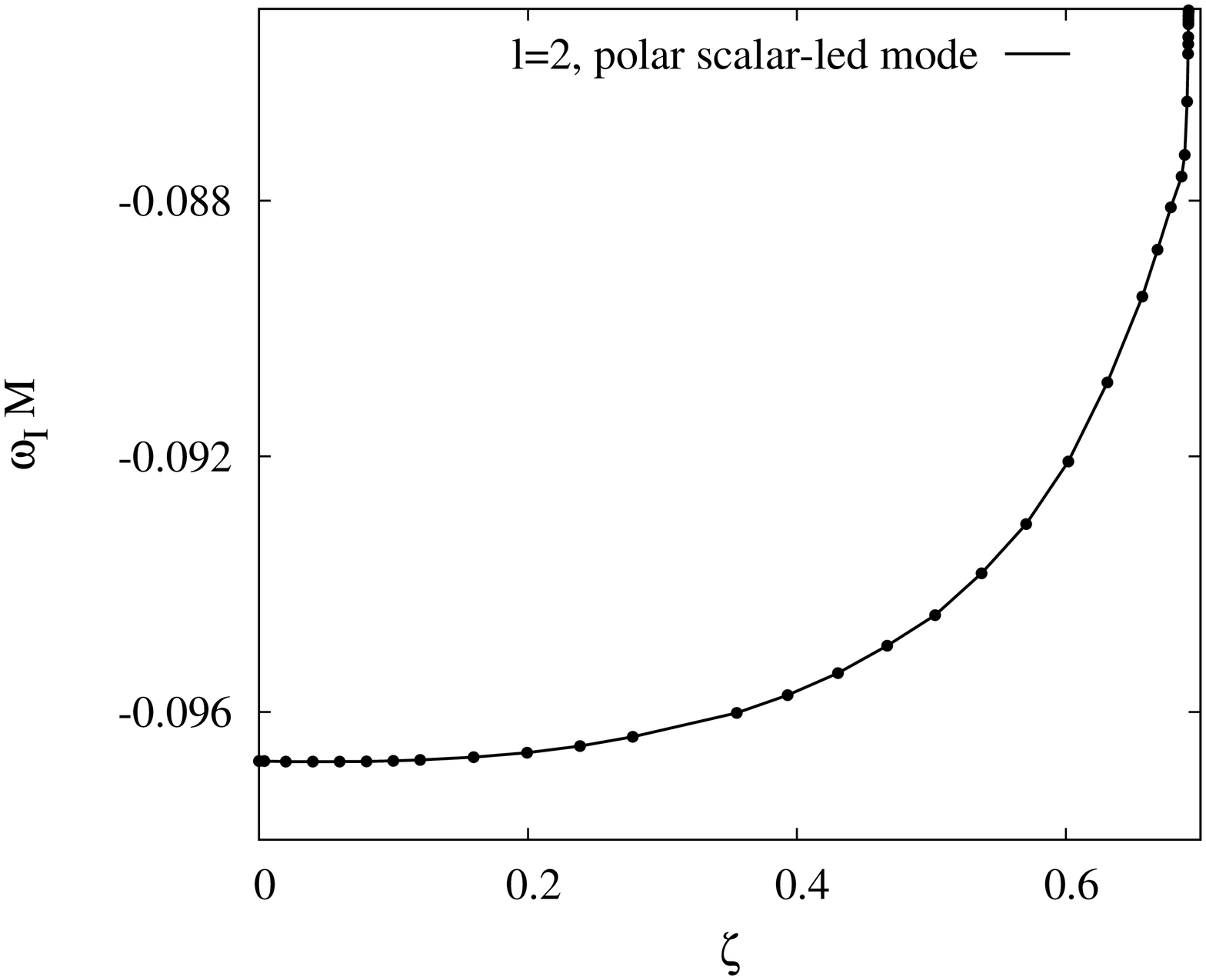}
         \caption{}
         \label{fig:dilaton_b}
     \end{subfigure}

     \caption{Scaled frequency $\omega M$
of the polar fundamental modes with $l=2$
as a function of the scaled GB coupling constant $\zeta = \alpha/M^2$.
(a) Real part $\omega_R M$ of the gravitational-led mode,
(b) imaginary part $\omega_I M$ of the gravitational-led mode,
(c) real part $\omega_R M$ of the scalar-led mode,
(d) imaginary part $\omega_I M$ of the scalar-led mode.}

         \label{fig:polar_l2}

\end{figure}

The polar perturbations include the excitation of the dilaton field, 
and hence we find two channels for the gravitational wave emission.
To classify these channels, we first recall that in GR
the gravitational $l=2$ frequencies are the same in
the axial and the polar case, i.e., there is
isospectrality \cite{Chandrasekhar:579245}.
The frequencies are $\omega M = 0.3737 - 0.08895i$ for $l=2$ 
and $\omega M = 0.5994 - 0.09270i$ for $l=3$.
But in addition there are also the modes of a scalar field in the
Schwarzschild background, 
$\omega M = 0.4836-0.09676i$ for $l=2$
and $\omega M = 0.6754-0.09649i$ for $l=3$.
Therefore, we refer to the modes arising 
in the limit $\zeta=0$ from the gravitational modes
as the gravitational-led modes, while we refer to the modes arising
in the limit $\zeta=0$ from scalar channel as the scalar-led modes.

In Fig.\ref{fig:polar_l2} we present the polar fundamental $l=2$ modes 
versus the scaled coupling constant $\zeta$. 
In general, the gravitational modes in the polar channels
show larger deviations from the known GR ($\zeta=0$) values,
as compared to their gravitational axial counterparts. 
In particular this means that the isospectrality 
of the axial and polar spectrum present for Schwarzschild black holes
is broken in EGBd theory. 

In Fig.\ref{fig:gravitational_a} we show the real part 
of the gravitational-led polar fundamental $l=2$ mode. 
As the GB coupling constant increases, the frequency decreases.
This is opposite to what happens for the axial mode 
shown in Fig.\ref{fig:l2_axial_a}. 
The reduction of $\omega_R M$ is, however, relatively small 
(below $2\%$).

The imaginary part $\omega_I M$ of the gravitational-led $l=2$ 
polar mode is shown in Fig.\ref{fig:gravitational_b} 
as a function of $\zeta$. 
As compared to GR, this mode tends to be more rapidly damped 
in EGBd black holes
for configurations below $\zeta=0.6$ (around a $2\%$).
However, for larger values of the coupling, 
the imaginary part decreases rapidly, 
resulting in more slowly damped modes than in the GR case. 
The structure close to $\zeta_L$ is more complicated, 
and we will comment more on it in Sec.\ref{2nd-branch}.

Let us now address the scalar-led modes. 
In Fig.\ref{fig:dilaton_a} we present the real part 
of the fundamental $l=2$ scalar-led mode as a function of $\zeta$, 
and in Fig.\ref{fig:dilaton_b} we present the corresponding imaginary part.

Fig.\ref{fig:dilaton_a} shows that 
for small values of the coupling constant, 
the real part of the frequency is slightly smaller 
for EGBd black holes
than for Schwarzschild black holes.
However, the frequency grows with the coupling from $\zeta=0.1$ onward. 
Around the limiting coupling $\zeta_L$ 
the frequency is about $8\%$ larger than in the GR case. 

The imaginary part decreases with the coupling, 
as seen in Fig.\ref{fig:dilaton_b}. 
Again, although for intermediate values of the coupling
the difference with respect to GR is not very large 
(up to $5\%$ around $\zeta=0.6$), 
close to the limiting coupling $\zeta_L$ 
the damping time can be much smaller than the GR case, 
as we will show in Sec.\ref{2nd-branch}.

\begin{figure}
     \centering

     \begin{subfigure}[b]{0.4\textwidth}
\includegraphics[width=50mm,scale=0.5,angle=-90]{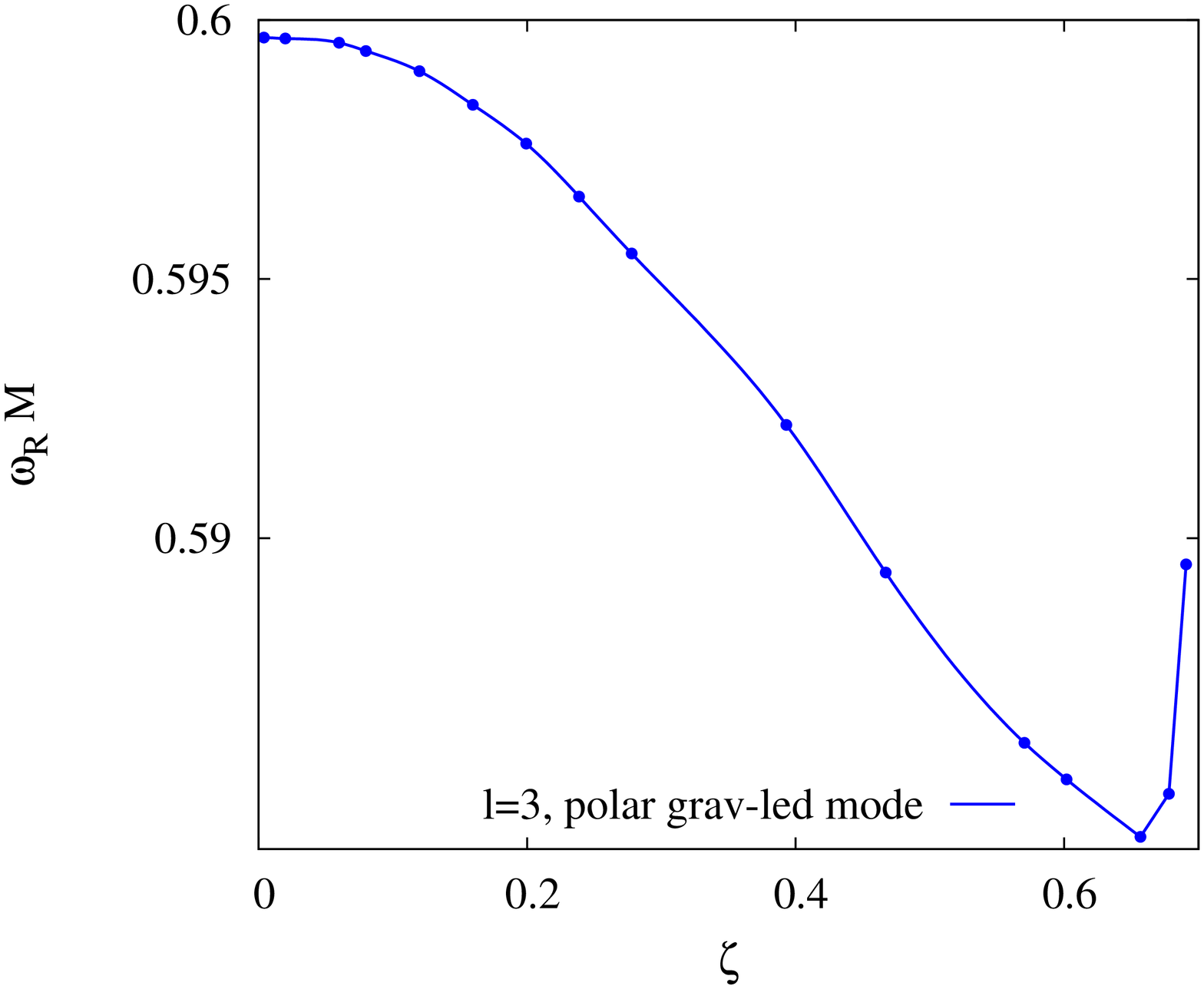}
         \caption{}
         \label{fig:gravitational_l3_a}
     \end{subfigure}
     \begin{subfigure}[b]{0.4\textwidth}
\includegraphics[width=50mm,scale=0.5,angle=-90]{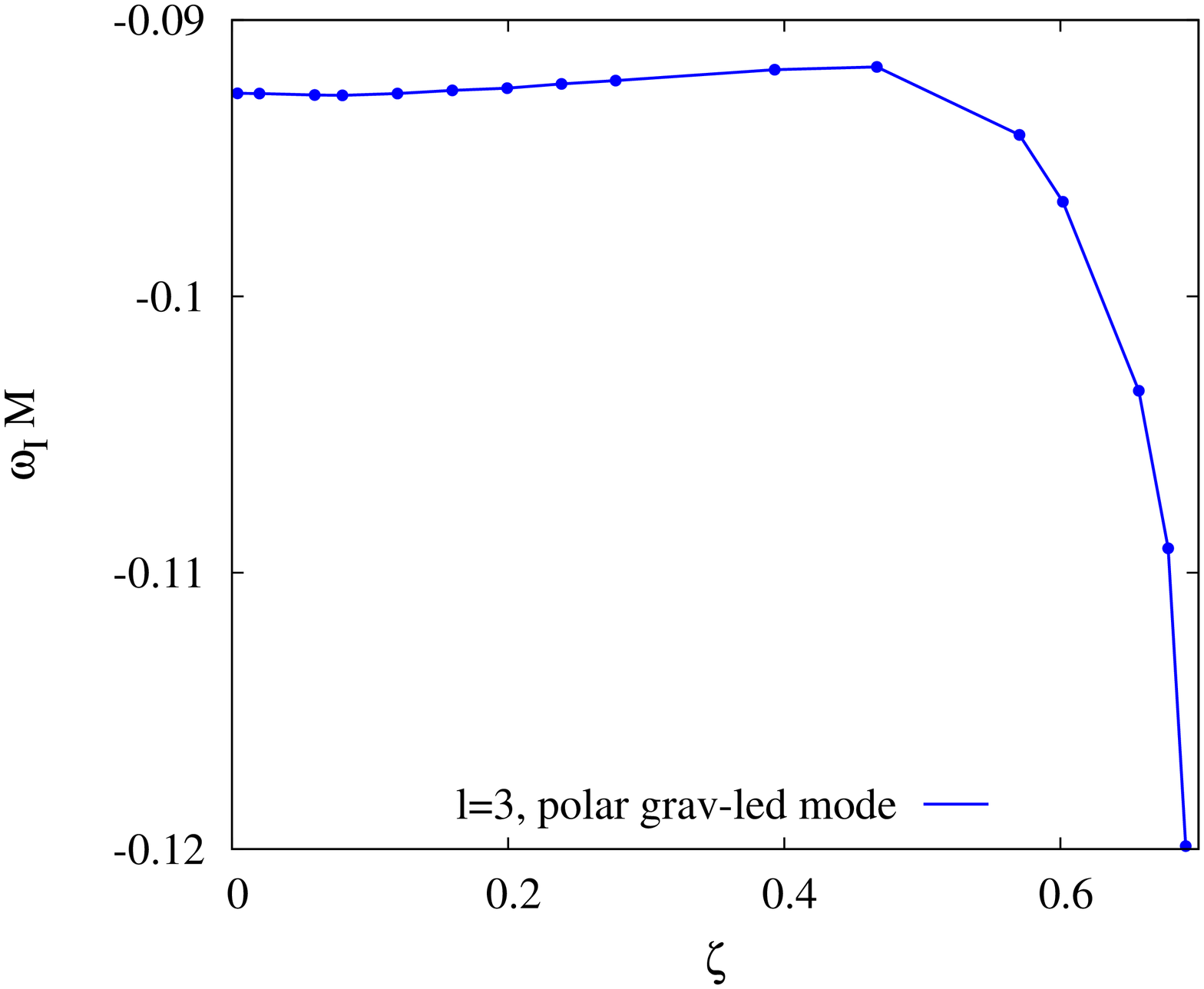}
         \caption{}
         \label{fig:gravitational_l3_b}
     \end{subfigure}
     \begin{subfigure}[b]{0.4\textwidth}
\includegraphics[width=50mm,scale=0.5,angle=-90]{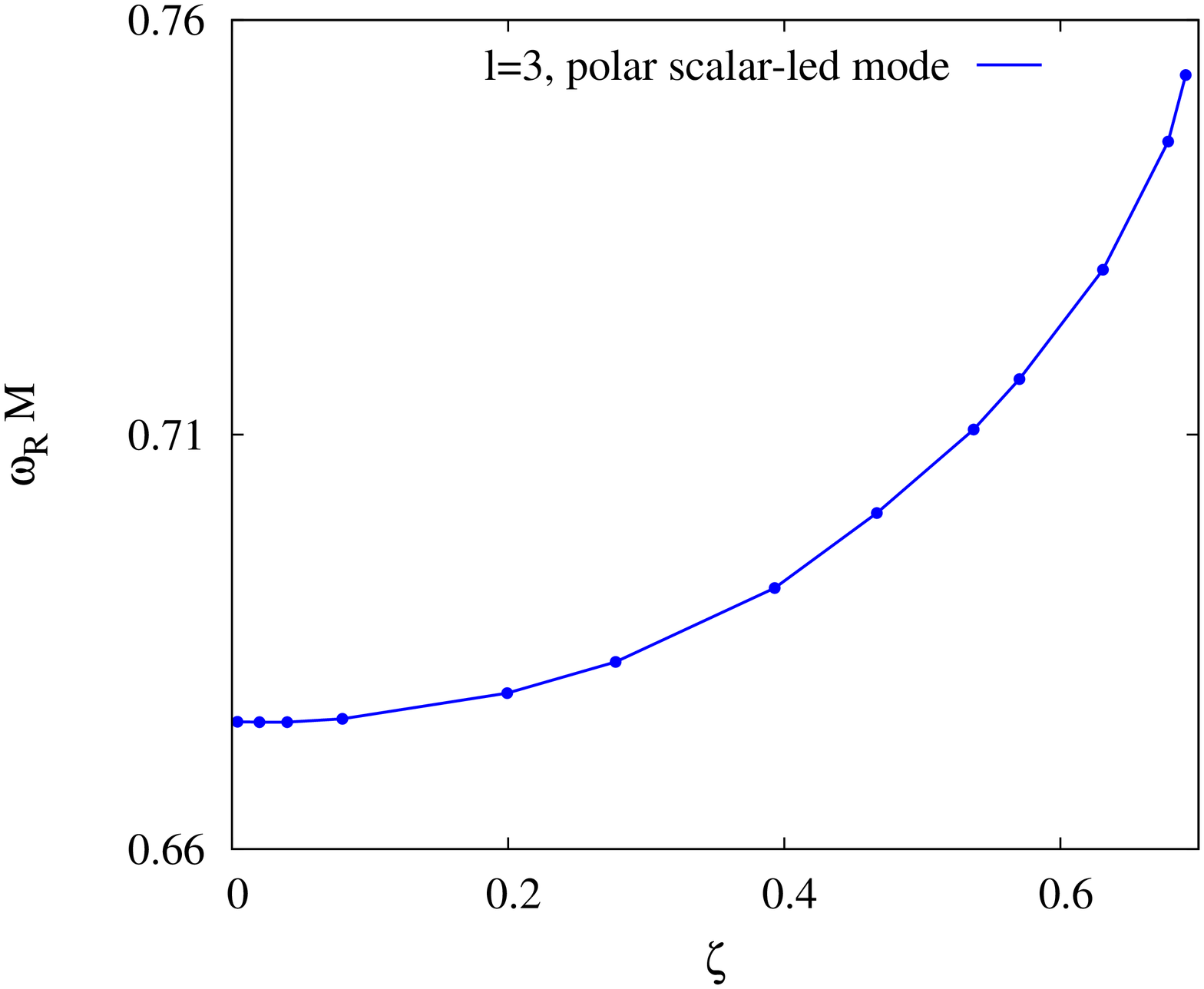}
         \caption{}
         \label{fig:dilaton_l3_a}
     \end{subfigure}
     \begin{subfigure}[b]{0.4\textwidth}
\includegraphics[width=50mm,scale=0.5,angle=-90]{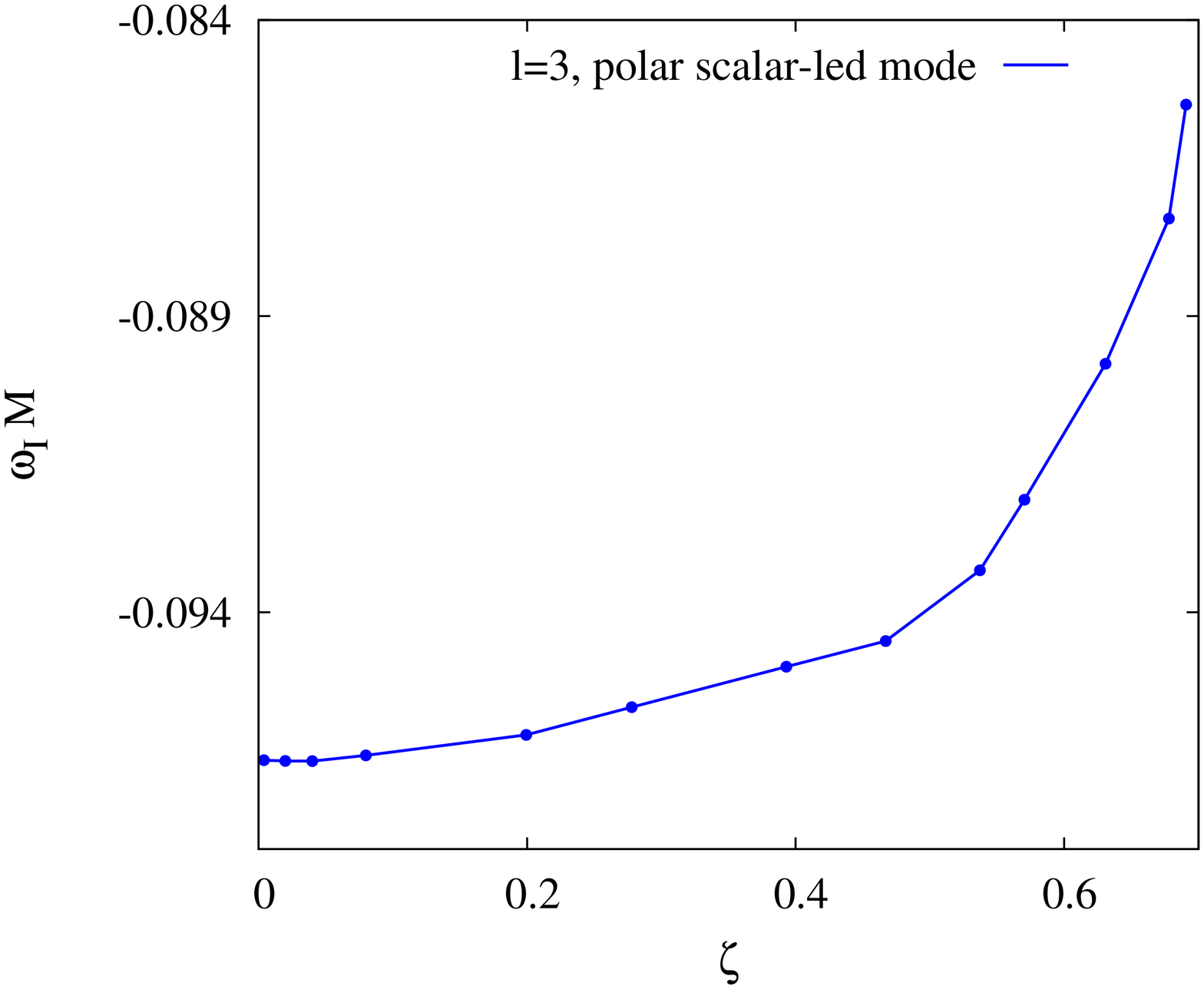}
         \caption{}
         \label{fig:dilaton_l3_b}
     \end{subfigure}
     
     \caption{Scaled frequency $\omega M$
of the polar fundamental modes with $l=3$
as a function of the scaled GB coupling constant $\zeta = \alpha/M^2$.
(a) Real part $\omega_R M$ of the gravitational-led mode,
(b) imaginary part $\omega_I M$ of the gravitational-led mode,
(c) real part $\omega_R M$ of the scalar-led mode,
(d) imaginary part $\omega_I M$ of the scalar-led mode.}

         \label{fig:polar_l3}

\end{figure}

Similar figures for the $l=3$ modes are shown 
in Fig.\ref{fig:polar_l3} for the gravitational-led and the scalar-led modes.  
The general behavior is similar to the $l=2$ modes. 
However, it is interesting to note that 
when increasing the coupling to $\zeta = 0.6$, 
the imaginary part of the $l=3$ gravitational-led mode 
increases significantly (Fig.\ref{fig:gravitational_l3_b}), 
meaning that the EGBd mode is more rapidly damped than its GR counterpart.

\subsection{Secondary branch instability}\label{2nd-branch}

In this section we study the quasinormal modes of the EGBd black holes 
in the vicinity of the maximal GB coupling $\zeta_L$ for dilaton
coupling $\gamma=1$.
At $\zeta_L$ the secondary branch of static EGBd black hole 
solutions appears, as discussed in Sec.\ref{sec_theory}. 

In \cite{PhysRevD.58.084004} an analysis of linear mode stability
was performed for purely radial perturbations.
This showed that the secondary branch is mode unstable.
In contrast, the primary branch does not present any such sign of instability.
This result is also in agreement with arguments from catastrophy theory
\cite{Torii:1996yi}.

\begin{figure}
     \centering

     \begin{subfigure}[b]{0.4\textwidth}
\includegraphics[width=50mm,scale=0.5,angle=-90]{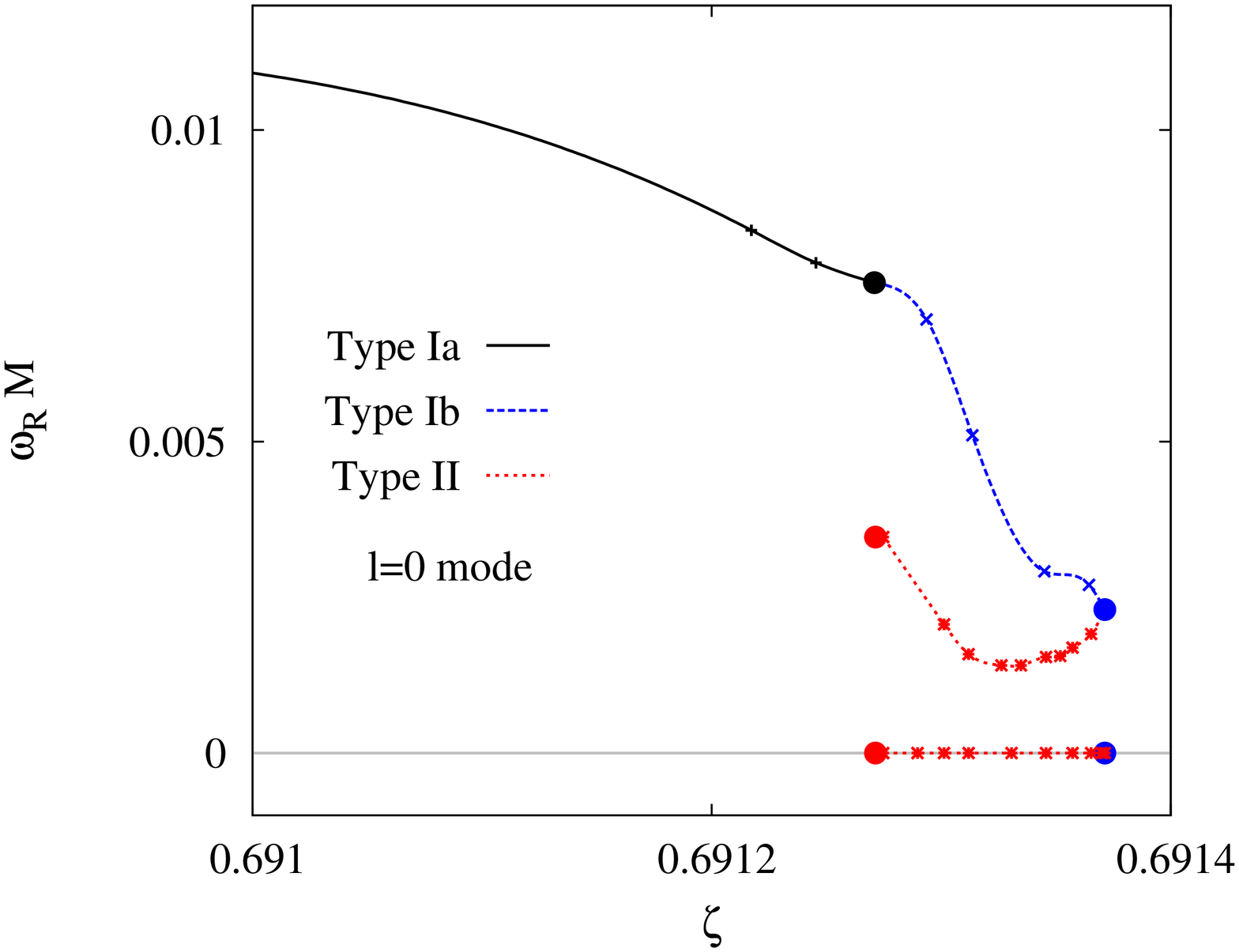}
         \caption{}
         \label{fig:l0_g_branches_a}
     \end{subfigure}
     \begin{subfigure}[b]{0.4\textwidth}
\includegraphics[width=50mm,scale=0.5,angle=-90]{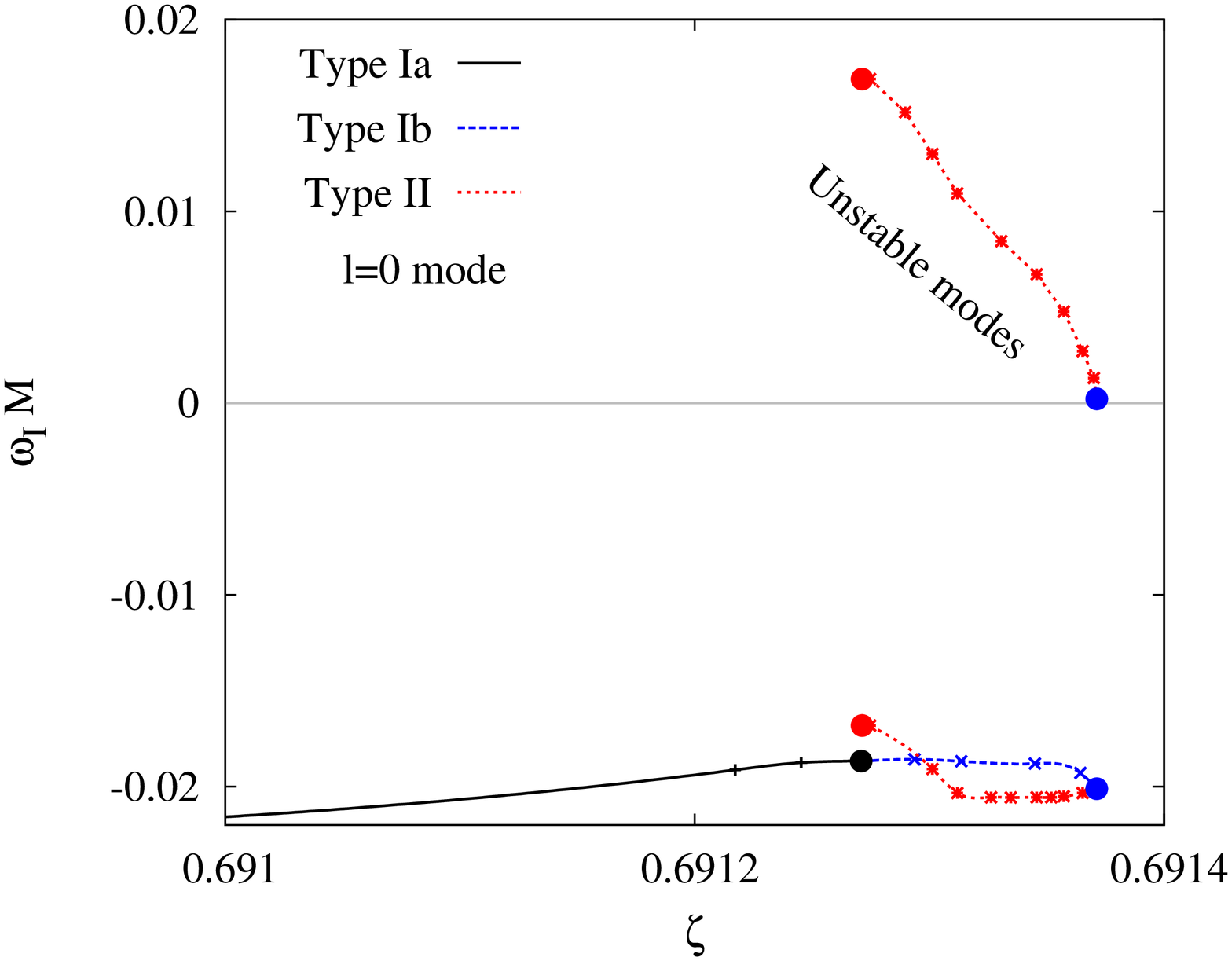}
         \caption{}
         \label{fig:l0_g_branches_b}
     \end{subfigure}
     
     \caption{Scaled frequency $\omega M$
of the $l=0$ modes as a function of the scaled GB coupling constant
$\zeta = \alpha/M^2$ close to its maximum value $\zeta_L$.
(a) Real part $\omega_R M$, (b) imaginary part $\omega_I M$.
Note that the half-plane with positive $\omega_I M$ represents unstable modes.}

         \label{fig:l0_g_branches}

\end{figure}

In Fig.\ref{fig:l0_g_branches} we exhibit a zoom into the region
of the GB coupling close to $\zeta_L$. Employing the same
color coding as in the figures of the solutions themselves,
we highlight the behavior of the previously discussed
$l=s=0$ mode in this region of the parameter space,
which connects with the Schwarzschild mode
in the limit $\zeta \rightarrow 0$.
These modes with negative imaginary part $\omega_I M$ 
are present for the type II as well as for the type I black holes.
However, close to $\zeta_L$ their imaginary part is much smaller than in 
the Schwarzschild case.

Interestingly, in addition to this branch of $l=0$ modes, which possesses
a Schwarzschild limit, there is a distinct further branch of $l=0$ modes,
which is present only for the solutions of the secondary branch.
This branch of modes represents the unstable radial modes
seen in \cite{PhysRevD.58.084004}.
These modes are purely imaginary, 
with the real part vanishing as seen in Fig.\ref{fig:l0_g_branches_a}. 
Fig.\ref{fig:l0_g_branches_b} shows
that imaginary part of these modes of the type II black holes 
resides in the positive half-plane. 
Thus these modes represent unstable modes. 
Considering the $\zeta \rightarrow \zeta_L$ limit on the secondary branch,
we note, that the imaginary part also vanishes there, 
and hence the mode disappears. 
Therefore these modes are not present on the primary branch 
of the type I black holes.

\begin{figure}
     \centering

     \begin{subfigure}[b]{0.4\textwidth}
\includegraphics[width=50mm,scale=0.5,angle=-90]{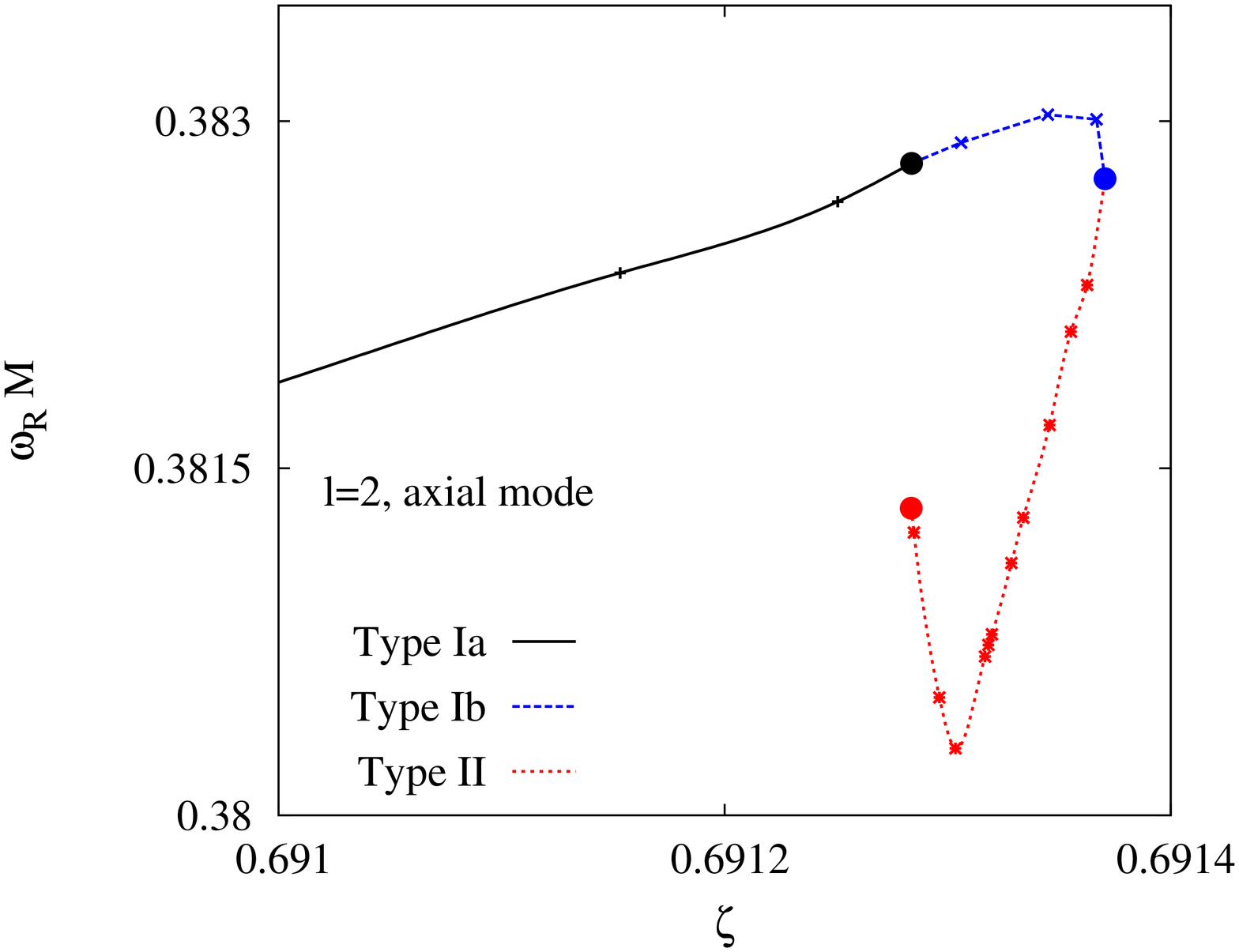}
         \caption{}
         \label{fig:l2_axial_branches_a}
     \end{subfigure}
     \begin{subfigure}[b]{0.4\textwidth}
\includegraphics[width=50mm,scale=0.5,angle=-90]{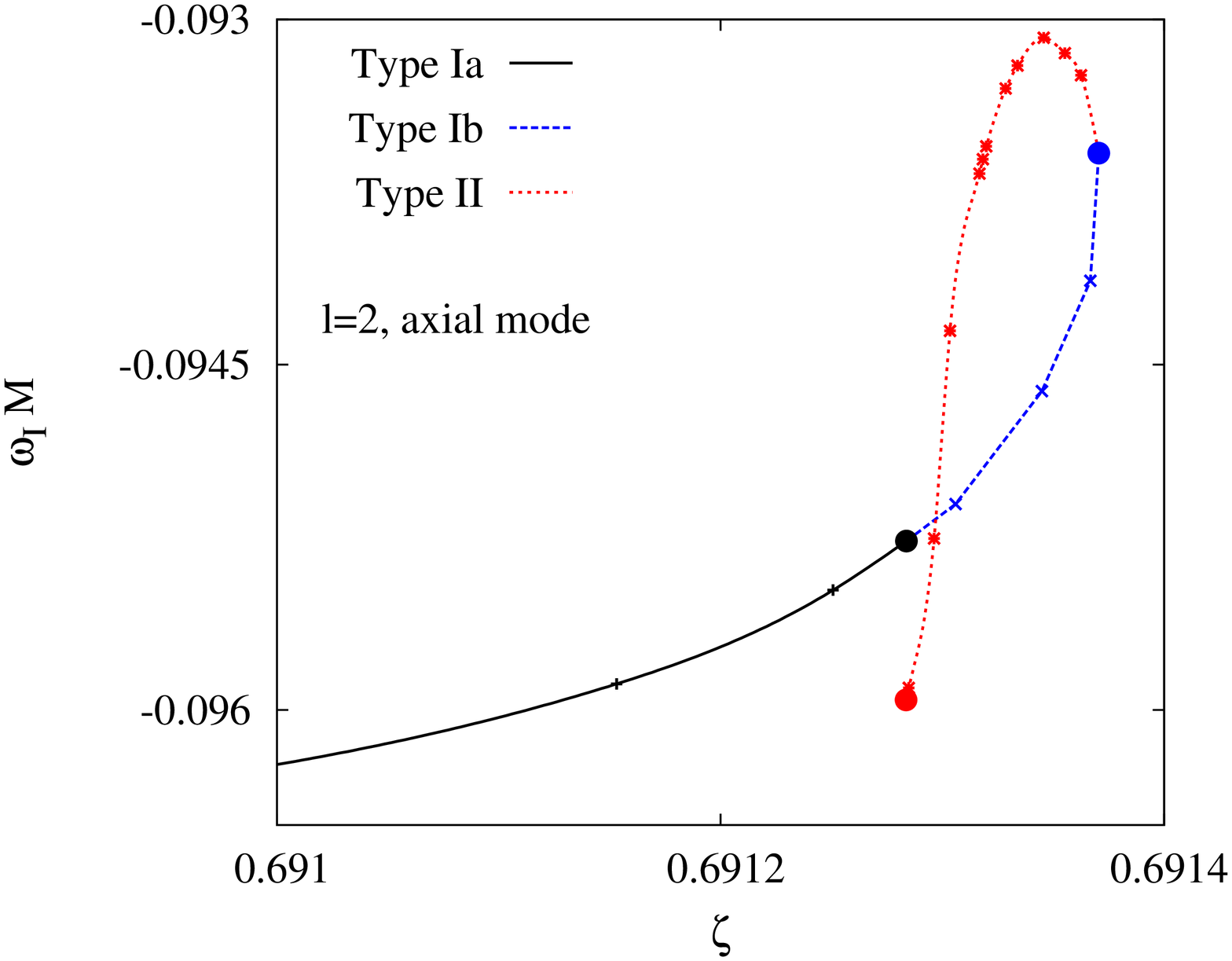}
         \caption{}
         \label{fig:l2_axial_branches_b}
     \end{subfigure}
     \begin{subfigure}[b]{0.4\textwidth}
\includegraphics[width=50mm,scale=0.5,angle=-90]{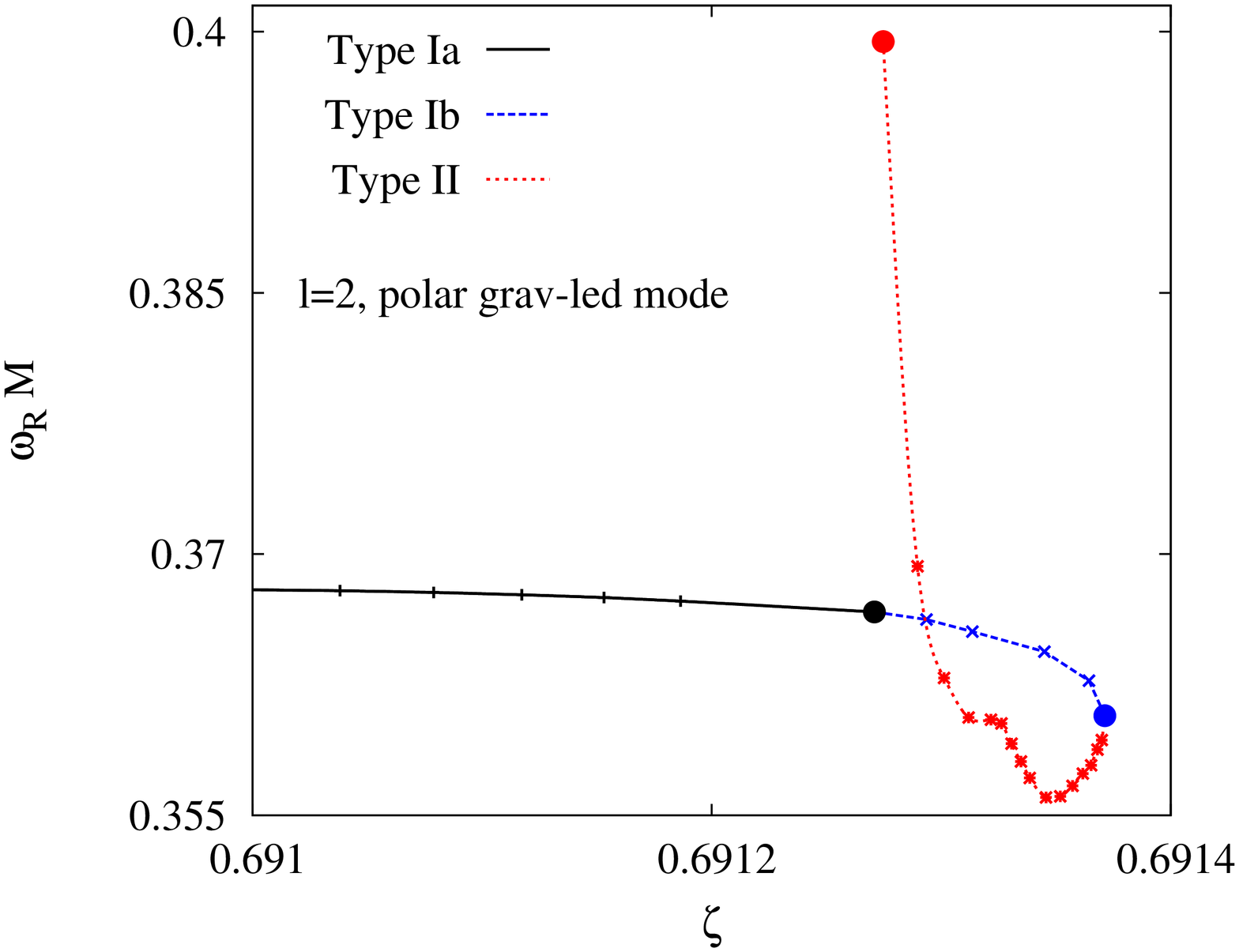}
         \caption{}
         \label{fig:l2_g_branches_a}
     \end{subfigure}
     \begin{subfigure}[b]{0.4\textwidth}
\includegraphics[width=50mm,scale=0.5,angle=-90]{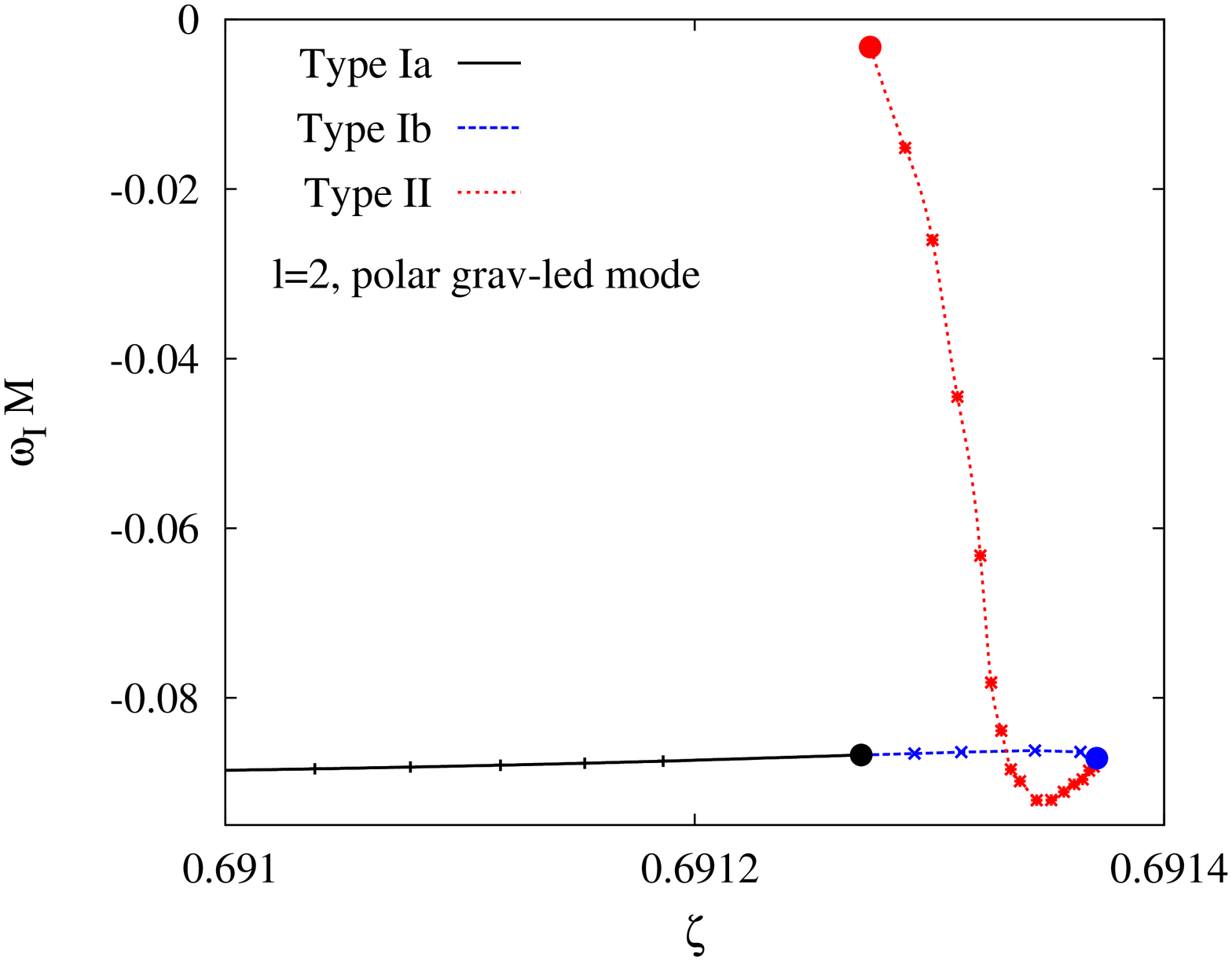}
         \caption{}
         \label{fig:l2_g_branches_b}
     \end{subfigure}
     \begin{subfigure}[b]{0.4\textwidth}
\includegraphics[width=50mm,scale=0.5,angle=-90]{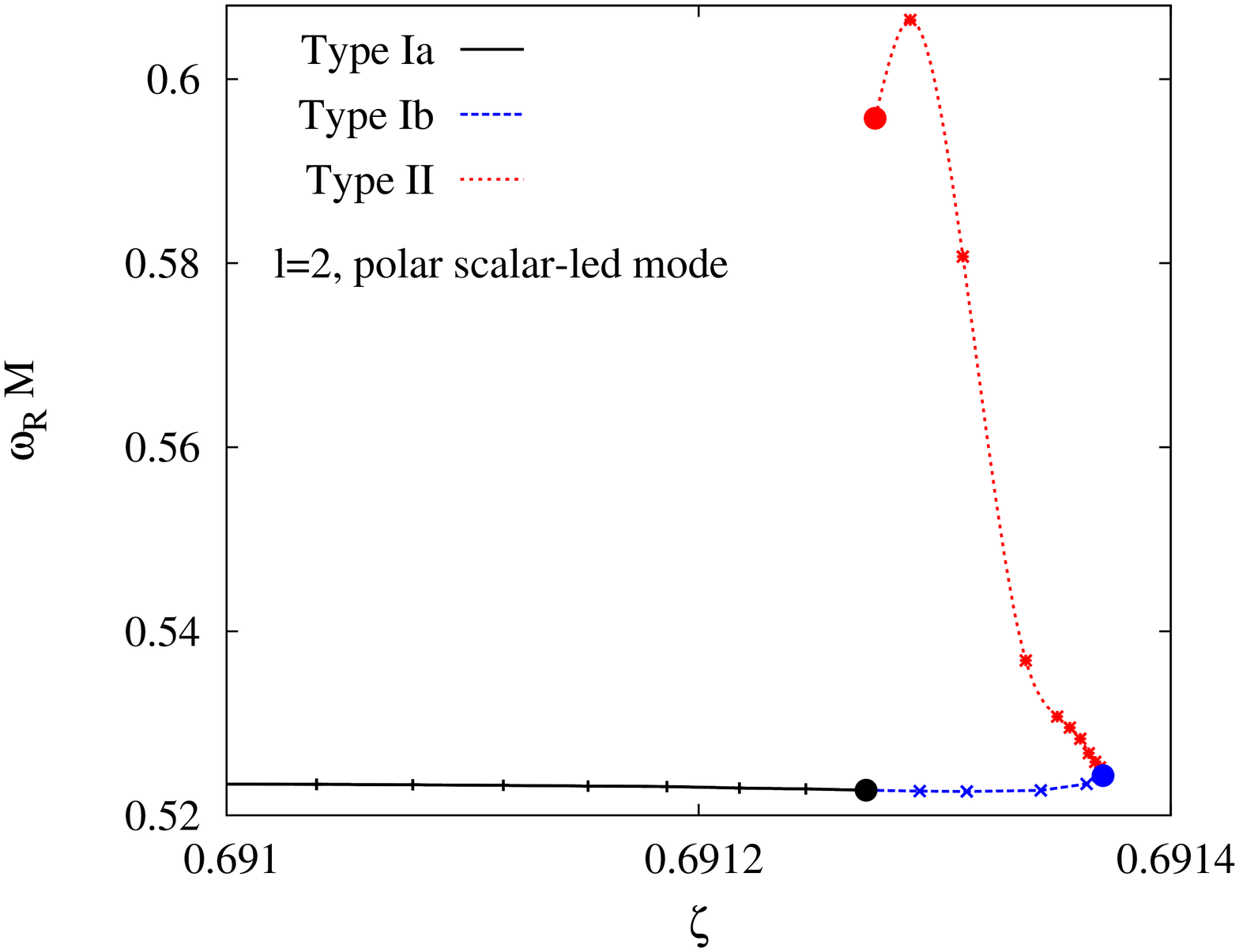}
         \caption{}
         \label{fig:l2_d_branches_a}
     \end{subfigure}
     \begin{subfigure}[b]{0.4\textwidth}
\includegraphics[width=50mm,scale=0.5,angle=-90]{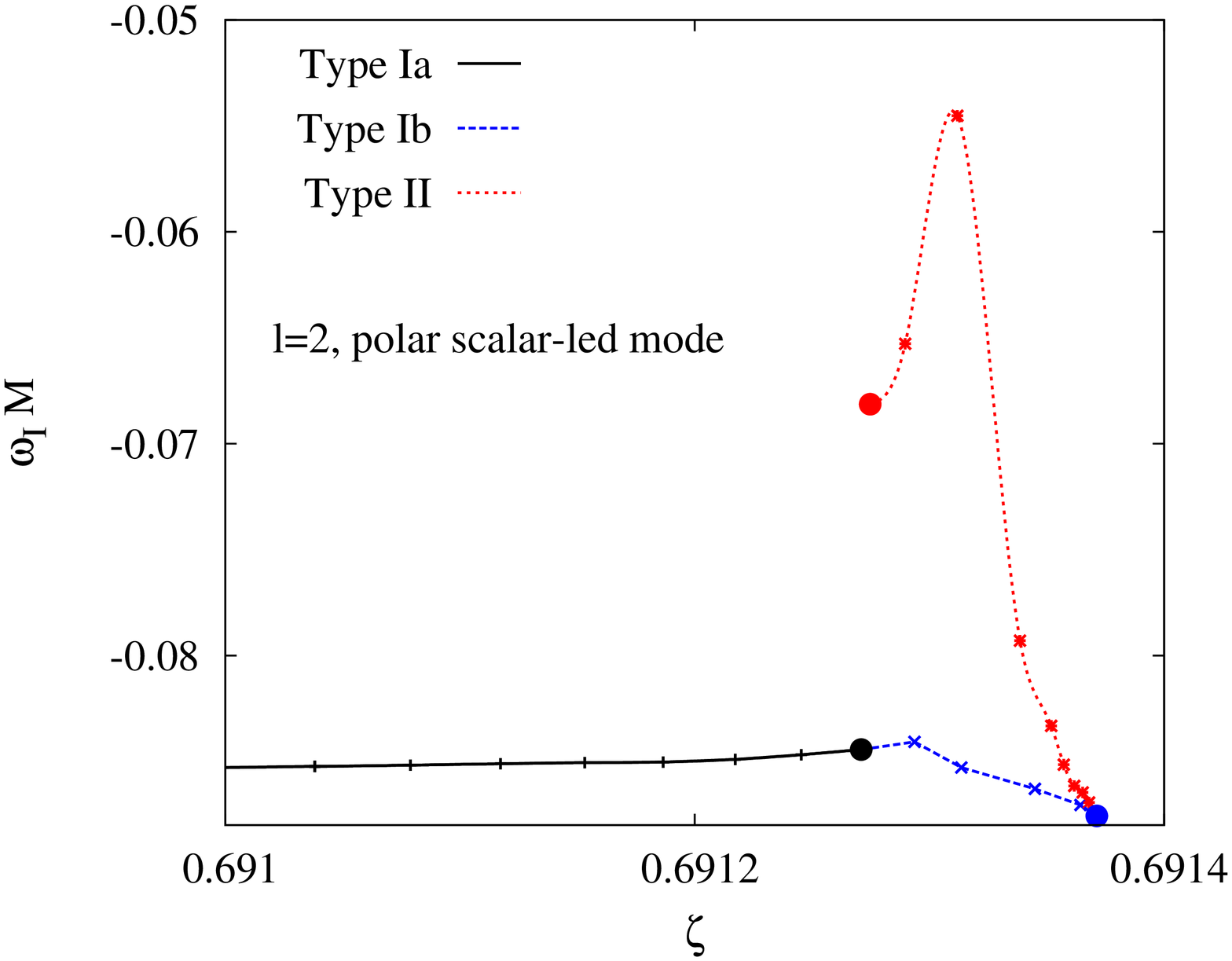}
         \caption{}
         \label{fig:l2_d_branches_b}
     \end{subfigure}

     \caption{Scaled frequency $\omega M$
of the fundamental modes with $l=2$
as a function of the scaled GB coupling constant $\zeta = \alpha/M^2$
close to its maximum value $\zeta_L$.
(a) Real part $\omega_R M$ of the axial mode,
(b) imaginary part $\omega_R I$ of the axial mode,
(c) real part $\omega_R M$ of the gravitational-led polar mode,
(d) imaginary part $\omega_I M$ of the gravitational-led polar mode,
(e) real part $\omega_R M$ of the scalar-led polar mode,
(f) imaginary part $\omega_I M$ of the scalar-led polar mode.}

         \label{fig:l2_branches}

\end{figure}

We exhibit in Fig.\ref{fig:l2_branches} 
the fundamental $l=2$ modes close to $\zeta_L$. 
The real part of the axial $l=2$ fundamental mode,
shown in Fig.\ref{fig:l2_axial_branches_a}, demonstrates
that type II black holes possess a lower frequency 
than type Ib black holes. 
The imaginary part displayed 
in Fig.\ref{fig:l2_axial_branches_b} 
is increasing with $\omega_I M$ for the type Ib black holes,
but decreasing again toward $\zeta_C$ for the
type II black holes.

The gravitational-led polar $l=2$ modes are exhibited in 
Fig.\ref{fig:l2_g_branches_a} (real part $\omega_R M$)
and Fig.\ref{fig:l2_g_branches_b} (imaginary part $\omega_R I$).
The frequency reaches a minimum at a particular type II configuration.
Close to $\zeta_L$ the frequency decreases continuously,
but then rises sharply to $\omega_{R} M=0.4$ at
$\zeta_C$.
The imaginary part shows a similar behavior,
rising up to $\omega_I M \approx -3 \cdot 10^{-3}$
when approaching $\zeta_C$.
Thus this mode of the type II black holes
is much longer ($\approx 30$ times) lived
than the corresponding modes of the type I solutions. 
Such a behavior is compatible with the beginning of an instability 
\cite{Brito:2015oca}.

The scalar-led polar $l=2$ mode,
shown in Fig.\ref{fig:l2_d_branches_a}
and Fig.\ref{fig:l2_d_branches_b},
exhibits only little $\zeta$ dependence on the
primary branch close to $\zeta_L$,
but a strong frequency increase of more than $10\%$
close to $\zeta_C$.
The imaginary part $\omega_I M$ features a cusp at $\zeta_L$ 
where the imaginary part reaches a minimum, 
followed by a strong increase for the type II solutions,
whose damping time can be $40 \%$ longer 
than for type I black holes. 
This effect in the scalar-led mode is qualitatively similar 
but less dramatic than in the gravitational-led mode.

In summary, 
we have not found any signs of unstable modes 
in type I black holes, even close to $\zeta_L$. 
The imaginary part of the frequency of the modes 
always remains negative in type I solutions.
However, we have confirmed previous results \cite{PhysRevD.58.084004}
showing that the $l=0$ modes of the black holes on 
the secondary branch, i.e., the type II black holes, are unstable. 
On the other hand,
by studying the $l=2$ modes of the secondary branch 
we have seen that in the gravitational-led polar $l=2$ modes, 
the damping time of the type II black holes 
could be $30$ times longer than in case of type I black holes,
if the instability in the $l=0$ channel could be avoided.

\subsection{Effect of the dilaton coupling $\gamma$ on the QNM spectrum}\label{sec_gamma}

\begin{figure}
     \centering

     \begin{subfigure}[b]{0.4\textwidth}
\includegraphics[width=50mm,scale=0.5,angle=-90]{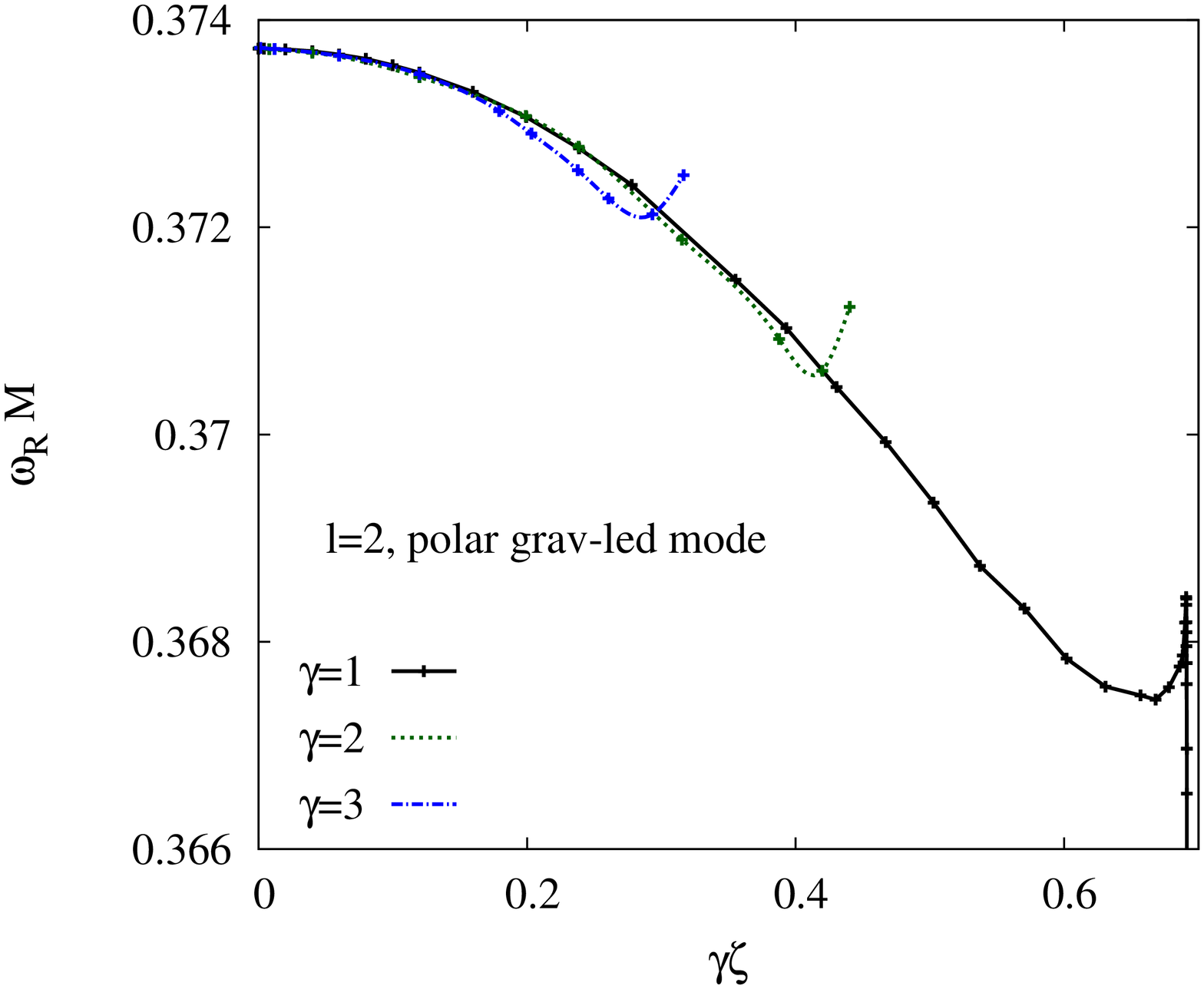}
         \caption{}
         \label{fig:l2_g_gamma_a}
     \end{subfigure}
     \begin{subfigure}[b]{0.4\textwidth}
\includegraphics[width=50mm,scale=0.5,angle=-90]{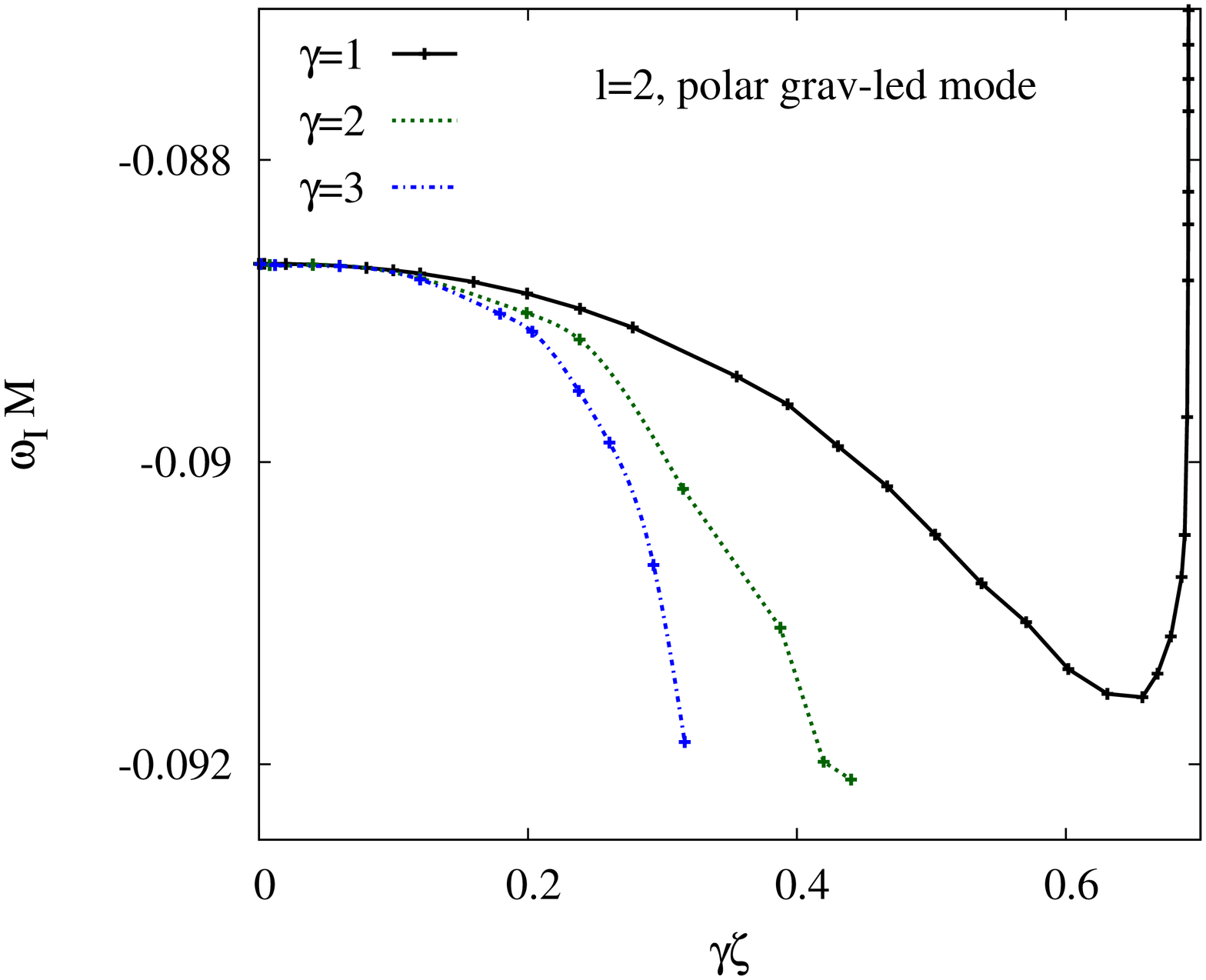}
         \caption{}
         \label{fig:l2_g_gamma_b}
     \end{subfigure}
     \begin{subfigure}[b]{0.4\textwidth}
\includegraphics[width=50mm,scale=0.5,angle=-90]{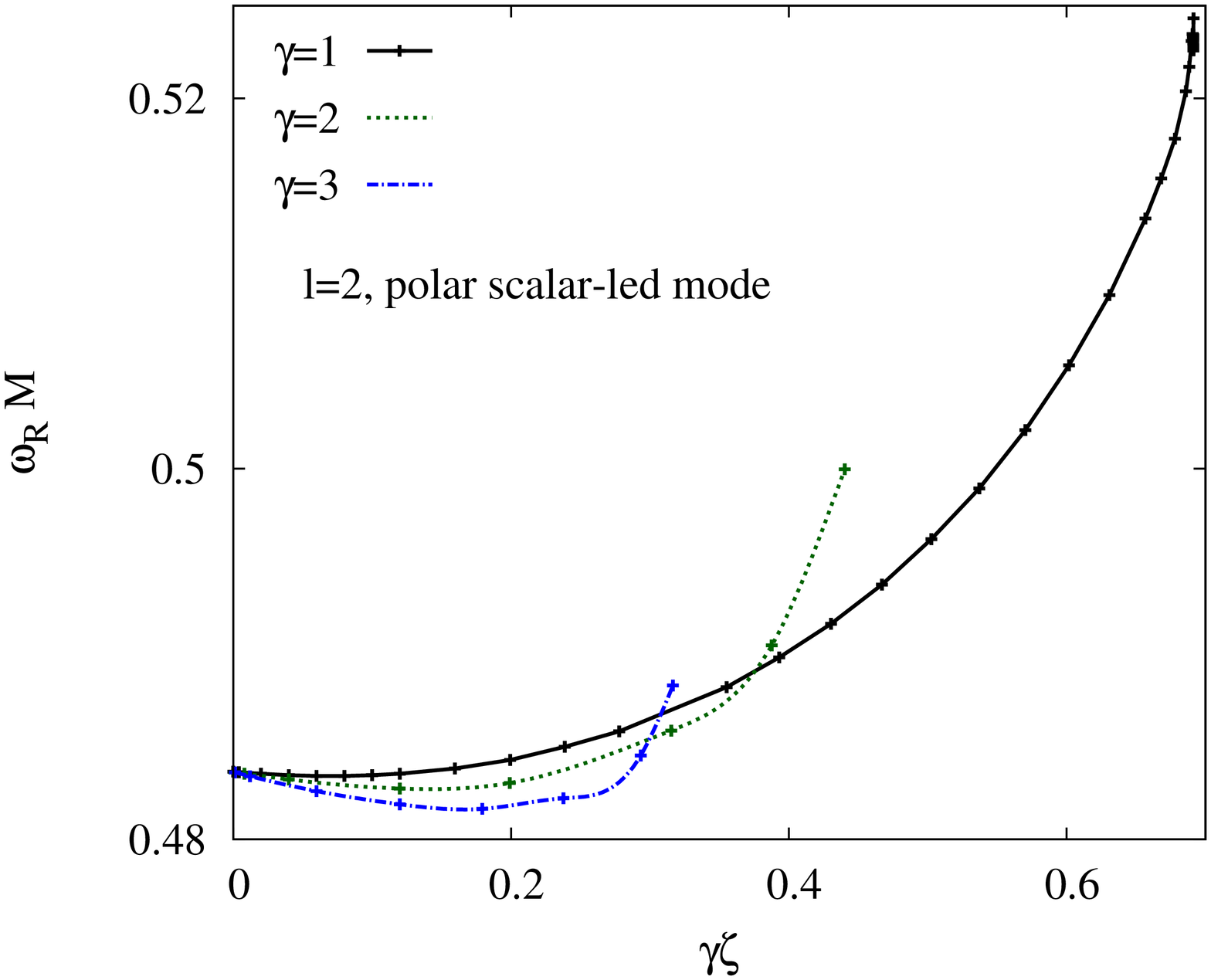}
         \caption{}
         \label{fig:l2_d_gamma_a}
     \end{subfigure}
     \begin{subfigure}[b]{0.4\textwidth}
\includegraphics[width=50mm,scale=0.5,angle=-90]{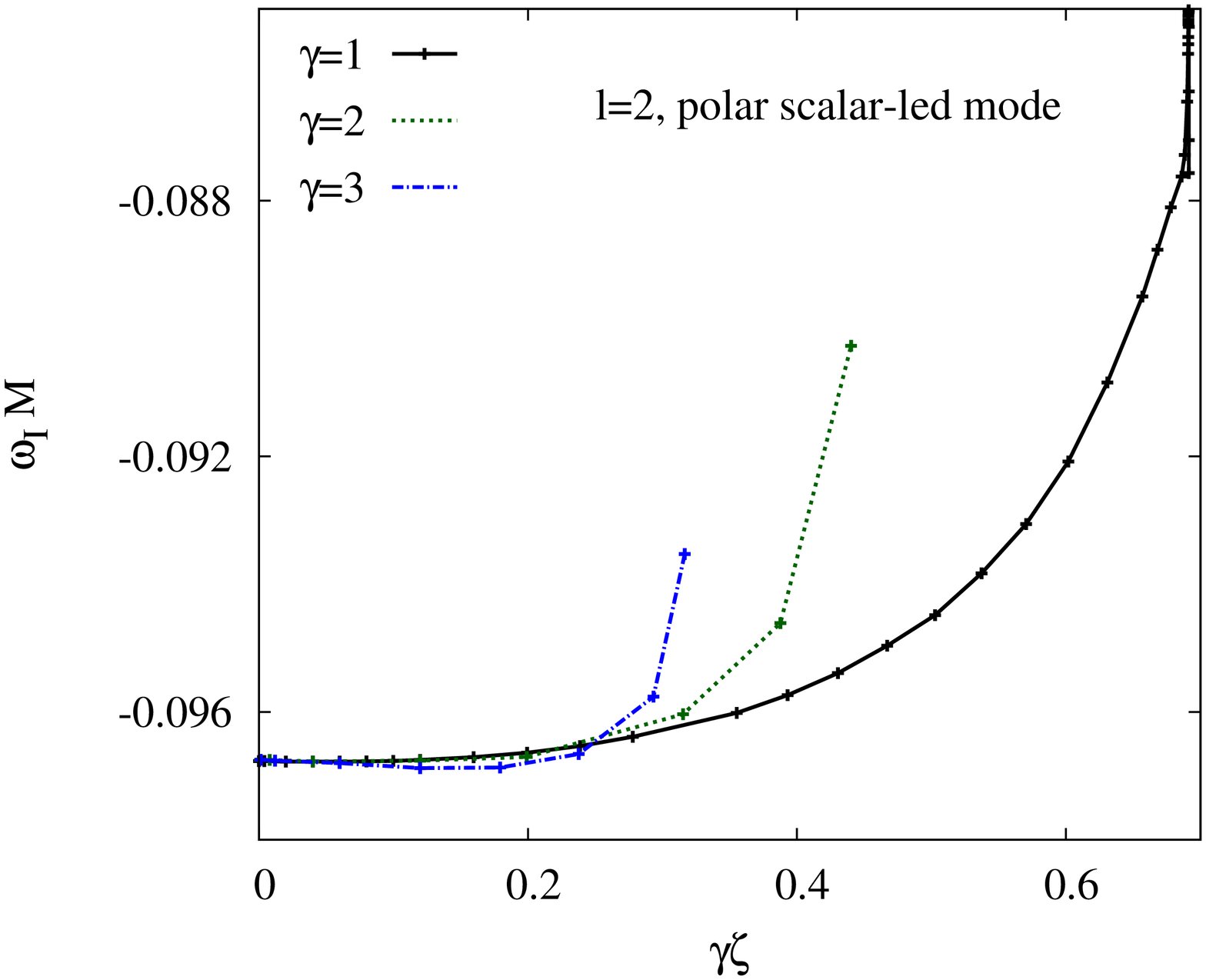}
         \caption{}
         \label{fig:l2_d_gamma_b}
     \end{subfigure}
     \caption{Scaled frequency $\omega M$
of the fundamental polar modes with $l=2$
as a function of the scaled GB coupling constant $\gamma\zeta$, with $\zeta = \alpha/M^2$
for dilaton coupling $\gamma=1, 2$, and 3.
(a) Real part $\omega_R M$ of the gravitational-led mode,
(b) imaginary part $\omega_I M$ of the gravitational-led mode,
(c) real part $\omega_R M$ of the scalar-led mode,
(d) imaginary part $\omega_I M$ of the scalar-led mode.}

         \label{fig:l2_gamma}

\end{figure}

So far we have considered the QNM spectrum
only for the case of dilaton coupling $\gamma=1$,
i.e., the value motivated by string theory. 
In this section we will address the effects for the
QNM spectrum obtained by varying the dilaton coupling $\gamma$.
In particular, we first focus on the fundamental $l=2$ polar modes
on the primary branch, and consider $\gamma=1, 2$ and $3$.

In Fig.\ref{fig:l2_gamma} we present the real and imaginary parts 
of the gravitational-led modes 
(Fig.\ref{fig:l2_g_gamma_a} and \ref{fig:l2_g_gamma_b}, respectively) 
and the real and imaginary parts of the scalar-led modes 
(Fig.\ref{fig:l2_d_gamma_a} and \ref{fig:l2_d_gamma_b}, respectively). 
As $\gamma$ is increased,
the frequencies change faster with $\zeta$,
but remain closer to the GR value, overall.
This also holds for the damping times of the scalar-led modes,
while the damping times of the gravitational-led modes
reach similar minimal values for all values of $\gamma$ considered.
We have employed $\gamma\alpha$ as the axis to demonstrate that the linear term dominates for a considerable range of values of the frequency, in particular, in the real part of the gravitational-led mode and the imaginary part of the scalar-led mode.

\begin{figure}
     \centering

\includegraphics[width=50mm,scale=0.5,angle=-90]{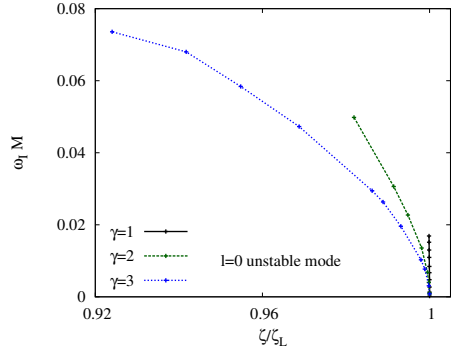}
     \caption{Imaginary part $\omega_I M$ of the scaled frequency
of the purely imaginary unstable $l=0$ modes
as a function of the normalized GB coupling constant $\zeta/\zeta_L$
for dilaton coupling $\gamma=1, 2$, and 3.
Note, that $\zeta_L=0.691372, 0.220144$ and $0.105532$
for $\gamma=1, 2$ and $3$, respectively.}

         \label{fig:l0_gamma}

\end{figure}

In Fig.\ref{fig:l0_gamma} we demonstrate the effect 
of the dilaton coupling on the instability of the black holes
on the secondary branch.
Since the real part of the corresponding $l=0$ unstable modes vanishes,
we exhibit only the imaginary part $\omega_I M$ versus the 
GB coupling.
However, we here employ a normalized GB coupling $\zeta/\zeta_L$, 
such that all three secondary branches start at the same point
$\zeta/\zeta_L=1$.
We note that the configurations obtained for larger values of $\gamma$ 
have a smaller $\omega_I $ for the same value of the mass $M$ 
and $\zeta/\zeta_L$ (within their domain of existence).
However, for the larger values of $\gamma$
the secondary branches extend further,
and the imaginary parts $\omega_I $ increase the more
toward $\zeta_C$ with larger $\gamma$.

\subsection{Astrophysical implications of the constraint on the Gauss-Bonnet coupling constant}

Here we comment on what we can learn from these results 
for the application to observations. 
We will focus on the case with $\gamma=1$, 
although a change of the dilaton coupling 
does not affect the qualitative features we will describe.

The current upper limit for the GB coupling constant $\alpha$, 
as mentioned already, 
derives from observations X-ray binary A0620-00 and is given by
$\sqrt{\alpha} \lesssim 3.8 \times 10^5$cm. 
Making use of the theoretical limit $\zeta_L=0.691372$
shows that, if we assume a theory with $\sqrt{\alpha}=3.8 \times 10^5$cm and $\gamma=1$,  
the smallest possible mass of a static black hole 
in the universe would be $M=3.095 M_{\odot}$.

On the other hand, the smallest stellar black hole candidate observed is in the X-ray transient GRO J0422+32. Originally, the central black hole of the binary system was thought to possess a mass of $M=4 \pm 1 M_{\odot}$ \cite{Gelino:2003pr}, but a more recent analysis of the data revealed a much smaller object with $2.1 M_{\odot}$ \cite{0004-637X-757-1-36}. 
This has cast doubts on the actual nature of the object in GRO J0422+32, which may not be a black hole, but another type of compact configuration. Interestingly, if this object is excluded, the data reveals a mass-gap between stellar black holes and neutron stars, with the minimum mass of black holes being compatible with $4.3 M_{\odot}$ \cite{0004-637X-757-1-36}. This mass-gap limit is for instance larger than the $M=3.095 M_{\odot}$ constraint imposed by A0620-00 on EGBd theory with $\sqrt{\alpha}=3.8 \times 10^5$cm and $\gamma=1$. 

The second LIGO/Virgo observation, GW151226 
\cite{PhysRevLett.116.241103} 
has detected gravitational waves from the coalescence 
of two stellar-mass black holes.
The comparison of the signal with GR simulations established 
that the masses of the black holes were
$M_{\text{BH}_1} = 7.5^{+2.3}_{-2.3} M_{\odot}$ 
for the first black hole BH$_1$, and
$M_{\text{BH}_2}=14.2^{+8.3}_{-3.7} M_{\odot}$ 
for the second black hole BH$_2$, 
which inspiraled and eventually merged. 
The resulting black hole BH$_3$ had an estimated mass of
$M_{\text{BH}_3}=20.8^{+6.1}_{-1.7} M_{\odot}$.
Since all these masses are larger than $M=3.1 M_{\odot}$, these observations do not improve the constraint A0620-00 on EGBd theory with $\gamma=1$, assuming they can be described by static EGBd black holes.

Note that if one makes the same reasoning assuming instead 
rotating black holes in EGBd, 
the value of the constraint will be smaller 
since the effect of rotation is to decrease $\zeta_L$ \cite{Kleihaus:2015aje}.
Observations appear to favour rotating black holes 
far from extremality \cite{PhysRevLett.116.241103}, 
hence the assumption of static initial black holes 
can be taken as a sufficient approximation. 
Similarly, one can consider only smaller GB couplings 
to allow for a lower black hole mass, 
since the analysis of X-ray transients \cite{0004-637X-757-1-36} 
suggests that the mass of stellar black holes can be equal 
or very close to the mass of large neutron stars ($2-3 M_{\odot}$).

\begin{figure}
     \centering

     \begin{subfigure}[b]{0.4\textwidth}
\includegraphics[width=50mm,scale=0.5,angle=-90]{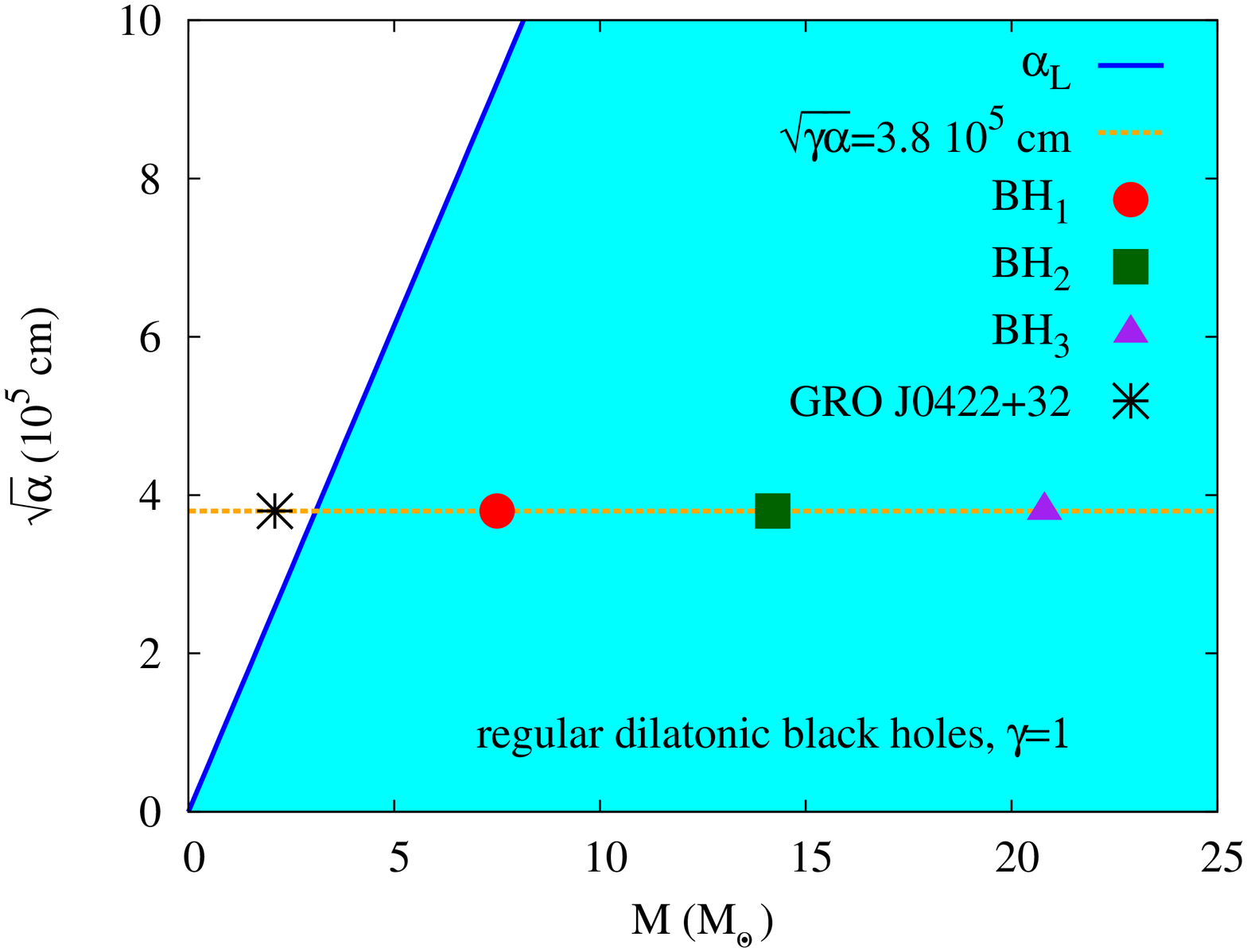}
         \caption{}
         \label{fig:alpha_M_eg_a}
     \end{subfigure}
     \begin{subfigure}[b]{0.4\textwidth}
\includegraphics[width=50mm,scale=0.5,angle=-90]{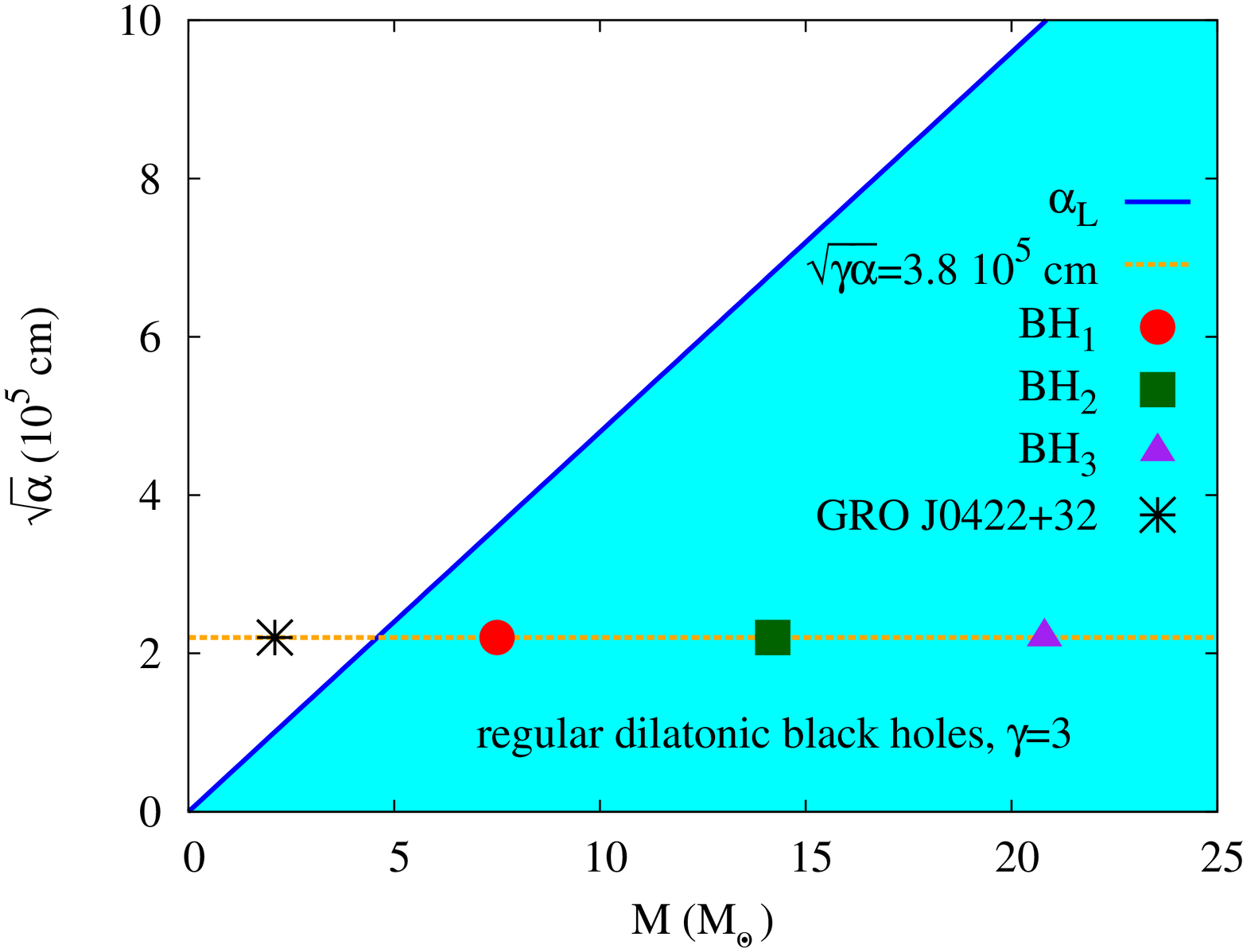}
         \caption{}
         \label{fig:alpha_M_eg_b}
     \end{subfigure}
     \caption{The square root of GB coupling constant $\alpha$ in cm 
versus the black hole mass $M$ in units of the solar mass $M_{\odot}$ for (a) $\gamma=1$ and (b) $\gamma=3$.
The blue line determines the limiting solutions, with $\alpha=\zeta_L M^2$.
The colored area below this line is the domain of existence of 
EGBd black holes.
The upper constraint on the coupling, $\sqrt{\gamma\alpha} = 3.8 \times 10^5$cm,
is marked with a dashed orange line.
On this line in red, green and purple we mark 
the black holes in the second LIGO/Virgo event,
BH$_1$, BH$_2$ and BH$_3$, and with a black asterisk we mark the black hole candidate GRO J0422+32.}

         \label{fig:alpha_M_eg}

\end{figure}

In Fig.\ref{fig:alpha_M_eg} we present the domain of existence 
of static black holes as a function of the mass in solar units, 
which is represented by the solid space in cyan. 
In blue we mark the limiting solutions with $\alpha=\zeta_L M^2$. 
In Fig.\ref{fig:alpha_M_eg_a} we show the domain for $\gamma=1$, and in Fig.\ref{fig:alpha_M_eg_b} for $\gamma=3$.
We mark the upper bound of the constraint 
on the coupling constant of $\sqrt{\alpha} = 3.8 \times 10^5$cm 
with an orange dashed line. On this line, in red, green and purple we mark 
BH$_1$, BH$_2$ and BH$_3$ respectively. With a black asterisk we mark the mass of GRO J0422+32. Note that this object is outside of the region of regular EGBd solutions, which implies that if this object were in fact a black hole, the constraint on the space of parameters would be improved considerably (for $\gamma=1$, $\sqrt{\alpha} = 2.57 \times 10^5$cm, and for $\gamma=3$, $\sqrt{\alpha} = 1 \times 10^5$cm).

\begin{figure}
     \centering

\includegraphics[width=50mm,scale=0.5,angle=-90]{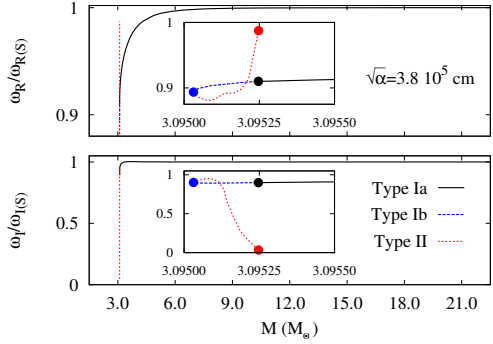}
     \caption{
Real and imaginary parts of the frequency $\omega$,
normalized with respect to the Schwarzschild frequency 
$\omega_{(\rm S)} = 0.3737 - 0.08895i$,
of the fundamental polar $l=2$ gravitational-led mode
versus the mass of black hole in units of solar mass $M_\odot$
for the GB coupling $\sqrt{\alpha}=3.8\times 10^5$cm.
The insets show the behavior close to 3.095$M_{\odot}$
(i.e., $\zeta_L$).}

         \label{fig:plot_ringdown_constraint4}

\end{figure}

We can ask the question of how does the QNM spectrum look 
in the case of such a coupling constant. 
For instance, consider Fig.\ref{fig:plot_ringdown_constraint4}, 
where we show the frequencies of the gravitational-led $l=2$ polar mode
normalized to corresponding Schwarzschild values, 
plotted as a function of the mass.  
Assuming the effective theory has indeed 
$\sqrt{\alpha} = 3.8 \times 10^5$cm, 
we show the deviation from GR of the real part (top) 
and imaginary part (bottom) as a function of the mass. 
The first thing to notice is that the final black hole 
with a mass around $21 M_{\odot}$ would present a QNM spectrum 
very similar to a GR black hole. 
The deviation in the ringdown frequency of the final BH$_3$ configuration 
with $21 M_{\odot}$ with respect to the Schwarzschild black hole 
is $0.07\%$, and $0.006\%$ in the imaginary part. 

Note, that this value of $\alpha$ supposes the most favorable scenario.
If one considers smaller values of $\alpha$ 
(hence allowing for an even smaller limit for the mass of black holes), 
the QNM spectrum in EGBd will be even closer to the GR description. 
Hence one can expect the ringdown phase of black holes 
far from $\zeta_L$ to be very well described by GR, 
which makes the measurement of $\alpha$ by
gravitational wave observations challenging. 
Of course in such a scenario, 
it would be possible to detect the dilaton component 
in the gravitational wave, 
although its quasinormal modes will be very close to the ones 
of a test scalar field. 
We conclude that the GR values would represent
a rather good approximation for the EGBd frequencies
in the description of the GW151226 event.

In order to constrain the theory further, 
it would be much more favorable to obtain observations 
of the ringdown of configurations close to $\zeta_L$,
since these would possess the largest deviations from GR 
in their spectrum. 
For instance, consider in Fig.\ref{fig:plot_ringdown_constraint4} 
solutions close to the maximal GB coupling $\zeta_L$ (blue dot). 
Type I black holes then exhibit deviations from GR on the order of $10\%$. 

While the fundamental gravitational wave frequencies of 
type II black holes deviate even stronger
from those of the Schwarzschild case, their $l=0$ radial instabilities 
should completely change the physical picture of the would-be
ringdown phase.

\section{Conclusions} \label{sec_conclusions}

We have continued the study of black hole quasinormal modes
in EGBd theory started in 
\cite{PhysRevD.79.084031,Blazquez-Salcedo:2016enn}.
Such quasinormal modes are thought to
describe the ringdown phase of an excited black hole, 
and can be detected by gravitational wave observatories 
such as LIGO/Virgo. 
The quasinormal modes contain the most direct and intrinsic 
information of a black hole, 
as from their detection the properties of the gravitational wave source 
can be inferred. 
Moreover, besides testing GR these observations can also
constrain alternative theories of gravity
such as EGBd theory.

Here we have further explored the linear mode stability 
of the static EGBd black holes. 
In particular, we have calculated the quasinormal modes 
from $l=0$ to $l=3$ in the full domain of existence
of the EGBd black holes.
Our results indicate the linear mode stability of the EGBd
black holes on their primary branch (type I black holes).
In agreement with previous results, 
we have found an instability in the $l=0$ spectrum 
of the black holes on the secondary branch (type II black holes), 
where the branch of unstable modes is purely imaginary.
While the $l=2$ modes of these type II black holes could be
very long-lived, this would require some mechanism
to evade the unstable $l=0$ modes.

The EGBd black holes break the isospectrality known for black holes in GR.
The black holes on the primary branch possess frequencies rather close
to the respective GR frequencies, unless they reside 
in the vicinity of the maximal GB coupling $\zeta_L$. This value
depends markedly on the value of the dilaton coupling $\gamma$.
In fact, we have considered the effect of variations of $\gamma$
on the QNM spectrum. 
The results show that the qualitative features of the $\gamma=1$ case 
are also present for other values of the coupling, 
in particular, we have observed a milder effect of $\gamma$ on the spectrum, 
and the continued presence of the instability on the second branch.

\begin{figure}
     \centering

\includegraphics[width=50mm,scale=0.5,angle=-90]{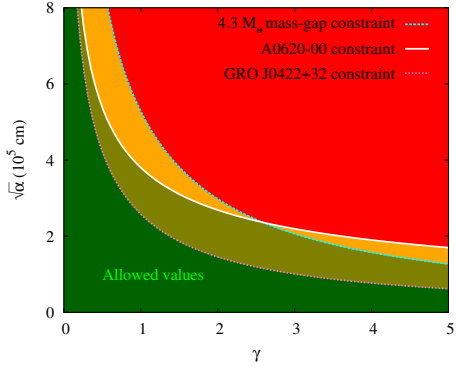}
     \caption{The parameter space of EGBd theory forms the plane ($\gamma$,$\sqrt{\alpha}$ ), in which we represent the current observational constraints: X-ray binary A0620-00 (solid white line) and mass-gap limit (dashed blue line). The red area represents the region of the parameter space excluded by both limits, and in orange by one of them. The green area represents the allowed values. The (possible) GRO J0422+32 black hole is shown as a dashed pink line, and can be used to further constrain the allowed region (light-green area).} 
         \label{fig:constraint_gamma}

\end{figure}

Our results suggest that constraining EGBd theory 
from the observation of the ringdown can be challenging. 
As an example, we have considered the black hole masses 
in the binary black hole coalescence GW151226,
invoking the theoretical limit on $\zeta$ (for different $\gamma$).
Considering the expected QNM frequencies have allowed
us to see that even in the most favorable case, 
the final black hole of $20.8 M_{\odot}$ 
would be ringing at a frequency very close to Schwarzschild. 
This suggests that in order 
to \textit{measure} or constrain the GB coupling constant 
from the detection of gravitational waves produced in
the ringdown phase of black hole mergers, one would need
black holes sufficiently close to the limiting GB coupling $\zeta_L$.

On the other hand, the observation of stellar black holes and the determination of their mass can be used to improve the constraint on the free parameters of EGBd theory, $\alpha$ and $\gamma$. In Fig.\ref{fig:constraint_gamma} we show the parameter space $\sqrt{\alpha}$ versus $\gamma$. The white line marks the constraint imposed by the X-ray binary A0620-00, and the blue dashed line the constraint imposed by the possible mass-gap. In red we mark the regions of the parameter space excluded by both limits, and in orange by one of them. The area in green represents values allowed by the observations. In addition we include the limit imposed by the possibility that GRO J0422+32 is a black hole with $2.1 M_{\odot}$, marked with a pink dashed line. In this case, both constraints on EGBd theory with arbitrary $\gamma$ are improved. The possible observation in the coming years of gravitational waves from small stellar mass black holes (if they exist) could be used to improve even more the current constraints of the parameter space of EGBd theory.

\section{Acknowledgments}

The authors thank Vitor Cardoso and Caio F. B. Macedo for helpful discussions. FSK thanks the University of Oldenburg and the University of Groningen for their kind hospitality. JLBS, FSK and JK gratefully acknowledge support by the DFG funded
Research Training Group 1620 ``Models of Gravity''. JLBS and JK gratefully acknowledge support from FP7, Marie Curie Actions, People, International Research Staff Exchange Scheme
(IRSES-606096).

\bibliographystyle{unsrt}

\bibliography{EGBdII_bib}

\end{document}